\documentclass[%
 reprint,
superscriptaddress,
 amsmath,amssymb,
 aps,
]{revtex4-2} 
\usepackage{amsmath}
\usepackage[ruled,vlined]{algorithm2e}
\usepackage{amsfonts}
\usepackage{bm}
\usepackage{xcolor}
\usepackage{verbatim}
\usepackage{graphicx}
\usepackage{siunitx}
\usepackage{upgreek}
\usepackage[normalem]{ulem} 
\usepackage{ tipa } 
\usepackage{caption} 
\usepackage{xr} 
\usepackage{gensymb} 
\usepackage{placeins}

\begin{document}

\raggedbottom
\author{Harry Penketh}
\email{h.penketh2@exeter.ac.uk}
\affiliation{Department of Physics and Astronomy, University of Exeter, Exeter, EX4 4QL, UK.}
\author{Sonal Saxena}
\affiliation{Department of Physics and Astronomy, University of Exeter, Exeter, EX4 4QL, UK.}
\author{Michal Mrnka}
\affiliation{Department of Physics and Astronomy, University of Exeter, Exeter, EX4 4QL, UK.}
\author{Cameron P.~Gallagher}
\affiliation{Department of Physics and Astronomy, University of Exeter, Exeter, EX4 4QL, UK.}
\author{Caitlin Lloyd}
\affiliation{Department of Physics and Astronomy, University of Exeter, Exeter, EX4 4QL, UK.}
\author{Diksha Garg}
\affiliation{Department of Physics and Astronomy, University of Exeter, Exeter, EX4 4QL, UK.}
\author{Christopher R.~Lawrence}
\affiliation{QinetiQ, Cody Technology Park, Ively Road, Farnborough, GU14 0LX, UK}
\author{Nicholas E.~Grant}
\affiliation{School of Engineering, University of Warwick, Coventry, CV4 7AL, UK}
\author{John D.~Murphy}
\affiliation{School of Engineering, University of Warwick, Coventry, CV4 7AL, UK}
\affiliation{School of Engineering, University of Birmingham, Edgbaston, Birmingham, B15 2TT, UK}
\author{David B.~Phillips}
\affiliation{Department of Physics and Astronomy, University of Exeter, Exeter, EX4 4QL, UK.}
\author{Ian R.~Hooper}
\affiliation{Department of Physics and Astronomy, University of Exeter, Exeter, EX4 4QL, UK.}
\author{Nick Stone}
\affiliation{Department of Physics and Astronomy, University of Exeter, Exeter, EX4 4QL, UK.}
\author{Euan Hendry}
\affiliation{Department of Physics and Astronomy, University of Exeter, Exeter, EX4 4QL, UK.}

\title{A microwave super-resolution imaging approach \\ towards breast cancer margin mapping}

\begin{abstract}
Accurate characterisation of margins in excised breast cancer tumours is critical to the success of surgical interventions. Yet margin status is typically confirmed post‑operatively using histopathology. Here we present a microwave single pixel imaging technique designed for use in intraoperative margin assessment. By leveraging the photo-induced change in microwave transparency of a silicon modulator placed under the sample, we map the microwave reflectivity of tissue-mimicking phantoms with deeply sub-wavelength resolution, allowing hydration mapping across large areas ($\sim$10~cm$~\times~10$~cm) at $\sim$1~mm resolution. We evaluate the discriminatory capability of our method using gelatin-based tumour phantoms with water-content variations designed to mimic the contrast between malignant tissue and tumour margins in resected breast specimens. We demonstrate the capability to identify, locate and quantify inadequate margins up to the typically targeted minimum thickness of 2~mm. Furthermore, using numerical modelling, we show that our approach is expected to be resilient to patient-specific tissue differences. These results establish microwave single-pixel imaging as a promising route towards real-time intraoperative assessment of margins in excised breast tumours.

\end{abstract}
\maketitle

\section{Introduction}
\begin{figure*}[htp]
    \centering
\includegraphics[width=0.95\textwidth]{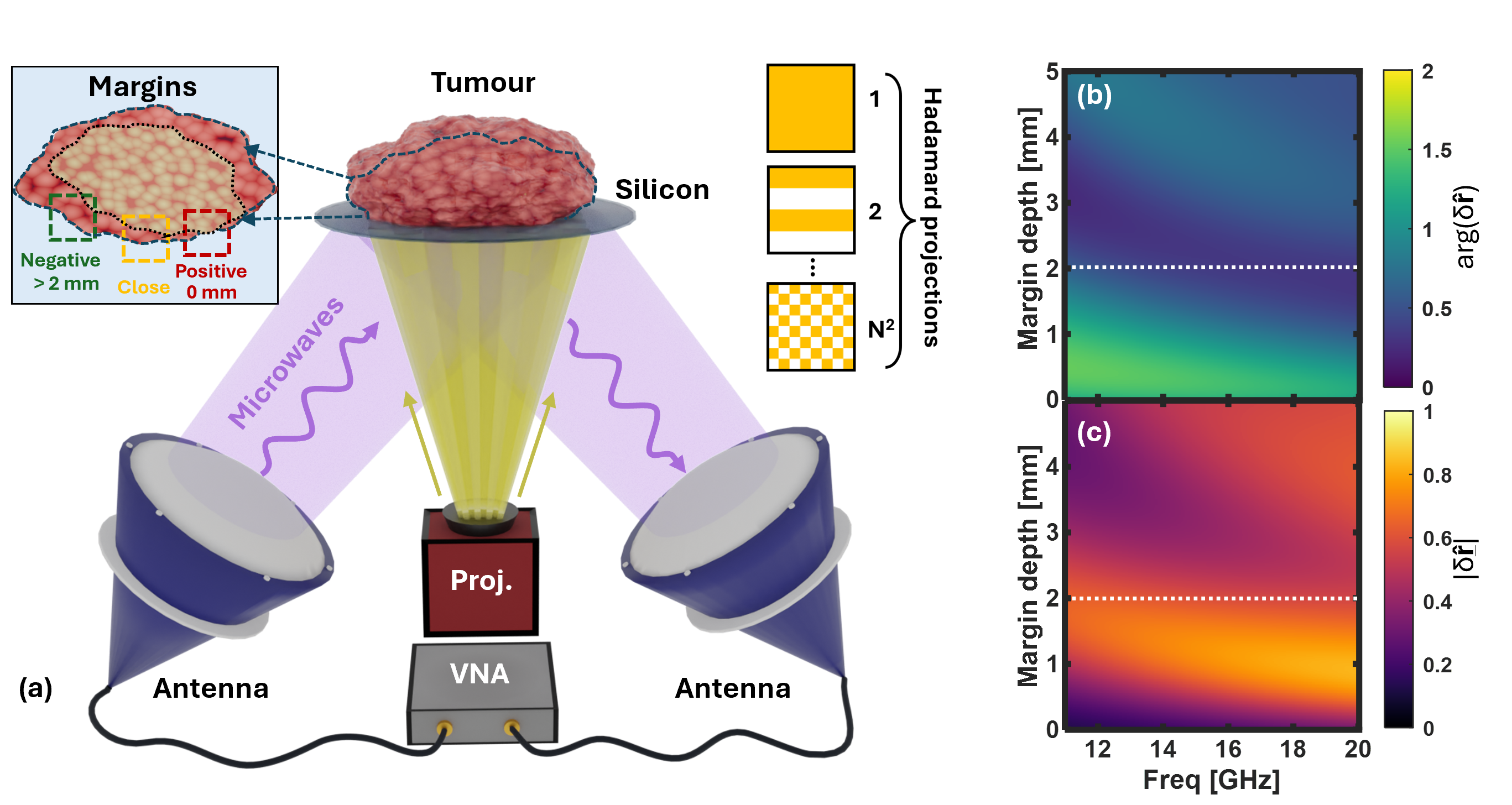} 
\caption{Method for margin depth mapping. (a) Experimental setup: a vector network analyser (VNA) records the photoinduced change in microwave reflection spectrum from a tumour phantom atop a silicon modulator. By projecting a sequence of binary Hadamard patterns (of $N~\times~N$ pixels - shown in right hand inset) and correlating each with a recorded spectrum, a hyperspectral image is formed. (Inset top left) lateral margins are classified as negative ($>$ 2~mm healthy tissue), close ($<$ 2~mm healthy tissue) and positive (no healthy tissue) as shown. Modelled changes in the normalised complex reflection $\hat{\delta\textbf{r}}$ when photoexcited for different margin depths are shown in (b) and (c) (phase and amplitude respectively). The UK target margin thickness of 2~mm is indicated by a dotted white line. 
    }
    \label{FIG:Setup}
\end{figure*}

Breast cancer is the most commonly diagnosed cancer in women worldwide, with around 2.3 million new cases and 670,000 related deaths each year \cite{Bray2024}. Breast cancer is mostly treated using breast conserving surgery (BCS). The objectives of BCS are twofold; the complete removal of the tumour including a margin of healthy tissue, whilst minimising the volume of healthy tissue removed. The presence of a cancer-free margin surrounding an excised tumour has been shown to reduce the chance of local disease recurrence significantly \cite{Houssami2014Margins}. Whilst margin guidelines vary internationally and with disease stage, the UK guidelines for breast cancer generally classify margins as positive (0~mm), close ($\textless$2~mm) and negative ($\geq$2~mm), as depicted in Fig. \ref{FIG:Setup} inset. Histopathology remains the `gold standard' for margin assessment in an excised tumour, but is typically carried out days after BCS. A $2017$ study revealed that in some UK units up to $40\%$ of patients undergoing BCS required reoperation to remove additional breast tissue, placing a significant burden on patients and healthcare providers \cite{Tang2017}. 

Intraoperative margin assessment (IMA) aims to identify inadequate margins during the initial surgery, thereby avoiding additional operations. However, there remains no clear leading approach which offers both high diagnostic accuracy and clinical feasibility \cite{schwarz2020,Heidkamp2021}. Typically, techniques that yield high resolution images such as optical coherence tomography \cite{Ha2018, Allen2018,Assayag2014,Chin2017}, microscopy \cite{Brachtel2016,Cahill2018,ChangTP,Tao2014,Vyas2017,Yoshitake2016} or Raman spectroscopy \cite{Wang2017Raman, Stone2021, Haskell2023} suffer from limited penetration depth through margins and/or long acquisition times, whilst approaches with excellent depth penetration such as X-ray computed tomography \cite{Tang2013,McClatchy2018,Qiu2018} and magnetic resonance imaging \cite{Golshan2014,Papa2016} have limited contrast and therefore low specificity.

Radiofrequency (RF) and terahertz electromagnetic radiation can discriminate cancerous from healthy tissues due to their differing hydration levels \cite{Chung_2008,Martellosio2015,Hussein2019,Hubbard2021}. However, high absorption by water impairs the use of terahertz radiation for evaluating deep breast cancer margins of millimetric thickness \cite{Yarina2022,Saxena2024}. RF margin assessment probes, which operate at lower frequencies, are commercially available, have been clinically tested,  and have proven successful in reducing secondary surgery rates \cite{Rossou2024,MarginProbeIFU2025}. However, RF probes have major limitations: they are limited to a binary discrimination between positive and negative margins, they have large sensing volumes ($\sim$ 7~mm~$\times~$7~mm \cite{MarginProbeIFU2025}) within which they are unable to localise small regions of positive margin, and they require labour intensive point-by-point measurements with limited sampling of the complete surface of the excised tumour.

An RF imaging approach to margin assessment could transcend these constraints. However, the long wavelengths of RF radiation severely restrict resolution in standard imaging. In this work, we overcome the spatial resolution problem of RF approaches to margin analysis using a single-pixel camera that exploits near-field photo-generated masks to enable deeply sub-wavelength resolution. Our single-pixel camera operates in the microwave portion of the RF band, specifically the frequency range 11 - 20 GHz which exhibits the maximal sensitivity to tissue depth. We experimentally demonstrate quantification of positive and close margins using tissue phantoms, up to the 2~mm limit appropriate for breast cancer margins \cite{ABS2024,Pilewskie2018}. In this proof of concept demonstration, a 100~mm~$\times$~100~mm area of excised tumour phantom is imaged hyperspectrally, and margin depth is algorithmically determined with $\sim$ 1 mm lateral resolution. Our approach offers a safe, low cost, sensitive method for intraoperative margin analysis, and potential to reduce the reoperation rate for breast cancer patients.

 \section{Results and discussion}
 \subsection{System design and simulations}
Our microwave single-pixel imaging system is shown schematically in Fig.~\ref{FIG:Setup}(a). It relies on the interaction of a series of optical projections with a microwave beam probing the sample under test. This interaction is facilitated by a silicon photomodulator, which converts incident visible photons into free charge carriers within the silicon, thereby locally blocking the microwave beam. In single-pixel imaging, discussed in detail in Method section~\ref{sec:imagingMethod}, an image is formed by correlating the sequence of measured signals (here the complex reflected electric field) with the known sequence of spatial masks, usually forming a Hadamard basis set \cite{Padgett2017,Gibson2020}. In our approach, as the microwave image resolution is governed by the dimensions of the pixels in the photoexcited masks, deeply subwavelength resolution is possible, with resolution of $10$s of microns demonstrated previously in this frequency band \cite{penketh2025}. 

The microwave probe beam is provided by a portable two port vector network analyser (VNA) coupled to two lens horn antennas in a reflection geometry, as shown in Fig.~\ref{FIG:Setup}(a). This gives collimated illumination of the photomodulator at 35$^{\circ}$ with transverse-electric (TE) polarisation, allowing convenient normal incidence photomodulation from a digital micromirror device (DMD). The compact VNA serves as both source and detector and monitors the frequency-dependent complex reflection coefficient $\mathbf{r}(f)$, capturing the change in both the amplitude and phase of reflected microwave radiation, over the range 11-20~GHz. Since the photomodulation induces a local change in reflection coefficient, our reconstructed images map the change in the complex reflection coefficient as a function of lateral position and frequency ($\delta \mathbf{r}(x,y,f)$). The value of $\delta\bf{r}$ depends on the illuminating microwave field and photoexcitation intensities, the photomodulator properties and crucially the presence of any material atop the modulator. To minimise susceptibility of the imaging to spatial inhomogeneity in the microwave and optical beams, experimental images and simulated data are normalised to the same signal in the absence of the sample ($\hat{\delta\bf{r}}$~=~$\delta\bf{r}_\text{SAM}/\delta\bf{r}_\text{REF}$, shown in supplementary Fig.~S7). 

We use the transfer matrix method to model the expected normalised signal $\hat{\delta\bf{r}}$ for a given margin depth and microwave frequency. In the transfer matrix method, described in detail in method section~\ref{sec:TMM}, the permittivities of the healthy and cancerous layers are assigned according to experimental characterisation of the tissue mimicking phantoms used for this work, also discussed in method section~\ref{sec:phantoms}. 

The breast tissue phantoms used in this work are composed of gelatine, water, safflower oil and kerosene in varying ratios following \cite{Lazebnik2005}, such that they can mimic both the fatty (adipose) margin tissue and the high water content glandular cancerous tissue (supplementary Fig.~S1) \cite{Lazebnik2007Meas}, with 20$\%$ water and 80$\%$ water, respectively. We vary the thickness of the margin layer, while the cancerous layer is $>$10~mm - this is thick enough so that transmission is zero. The permittivities (and therefore losses) in our fabricated margin phantoms are higher than some reported literature values for breast tissues (see supplementary Fig.~S1(c) and (f)), with a 20$\%$ water content margin layer lying at the upper end of literature values for healthy breast tissue \cite{Lazebnik2007MeasCancer,Kuwahara2020}. This was selected as the most challenging scenario expected to occur in real tumours, with enhanced imaging sensitivity and depth predicted for real patients (as shown in supplementary Fig.~S1).

 The phase and amplitude of $\hat{\delta\mathbf{r}}$ predicted by the transfer matrix method for different margin thicknesses is shown in Figs.~\ref{FIG:Setup}(b) and (c). We choose this frequency range as it corresponds to a quarter wavelength Fabry-P\'erot condition within close margins, which gives strongly varying reflection as a function of both frequency and margin depth. This makes our measurement the most sensitive to small changes in tissue thickness for margin depths in the range 0 to 2~mm - exactly what one would want for breast cancer margin analysis. Higher order modes are also apparent for greater margin depth and/or higher frequency, which exhibit reduced contrast due to propagation losses through the margin tissue. This degeneracy necessitates a complex (amplitude and phase) or spectral measurement for depth disambiguation. By measuring a complex spectrum, we reduce susceptibility to experimental artefacts and noise.

\subsection{Margin depth determination}
\begin{figure}[tp!]
    \centering
\includegraphics[width=0.45\textwidth]{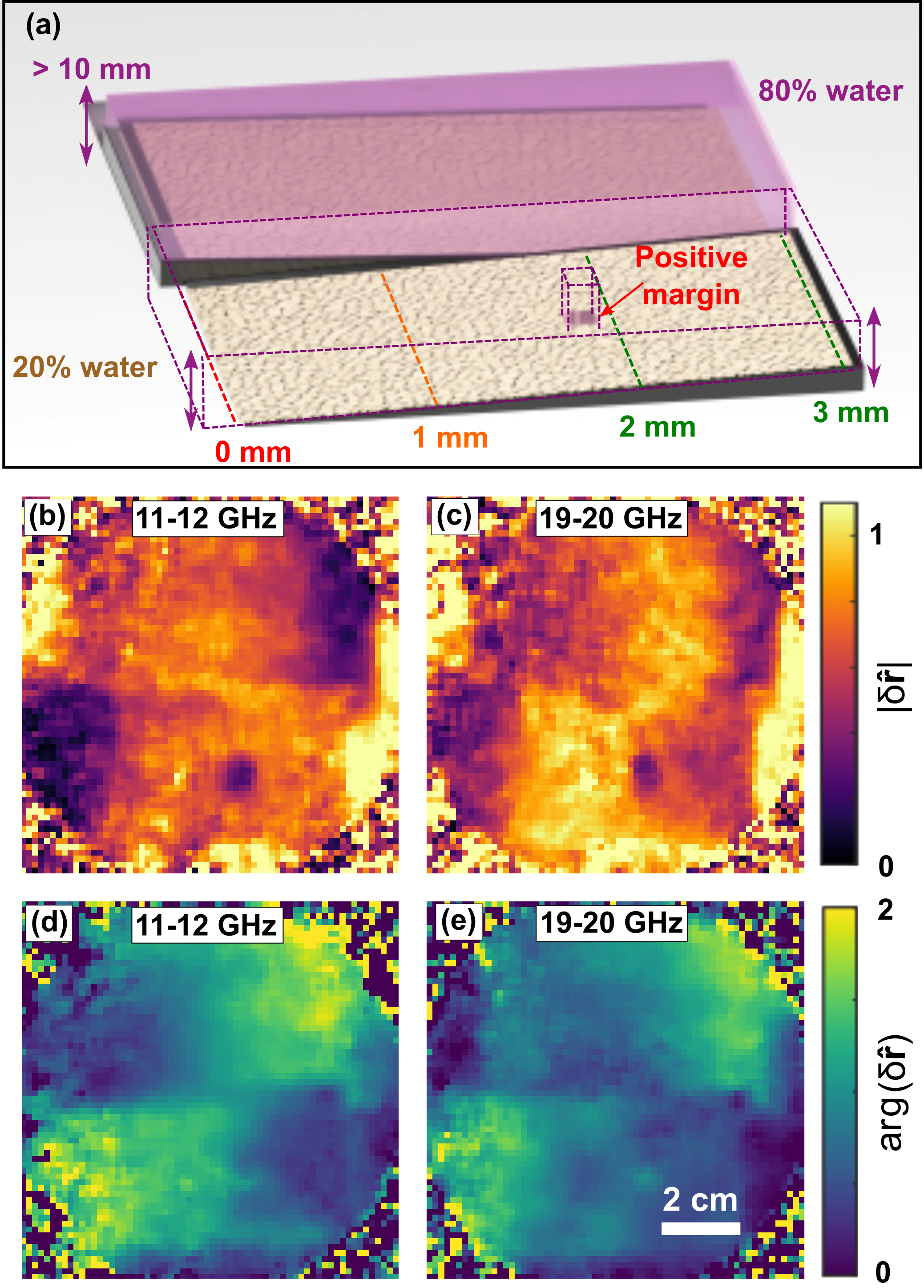} 
    \caption{Single-pixel imaging of a dual ramp breast tumour phantom. (a) Schematic of the layered tissue phantom mimicking breast cancer margins, which comprises two counter-inclined linear ramps of healthy tissue phantom (20$\%$ water content), topped with thick cancerous tissue phantom (80$\%$ water content). On the lower ramp, a small positive margin is substituted into the healthy phantom layer. (b) and (c) $64\times64$ pixel images of the photoinduced change in reflected microwave intensity $|\hat{\delta\mathbf{r}}|$ of the layered phantom in (a), averaged over 11-12~GHz and 19-20 GHz spectral regions respectively. (d) and (e) as in (b) and (c) but for phase arg($\hat{\delta\mathbf{r}}$).}
    \label{FIG:results}
\end{figure}

To provide a known spatially varying margin depth, we produce a bi-layer gelatine phantom which comprises two counter-inclined linear ramps of margin thickness, shown in Fig.~\ref{FIG:results}(a). The 20$\%$ water layer is cast into a 3D printed acrylonitrile butadiene styrene (ABS) plastic mould with thickness ranging from 0~mm to 3~mm. The $80\%$ water overlayer is positioned on top of each ramp. In addition, we include a localised cancerous protrusion into the thick margin layer of one of the ramps as shown, forming an isolated positive margin of lateral size $\sim$5~mm~$\times$~8~mm.
\indent

Our experimental image reconstructions form a complex valued, hyperspectral data cube, with dimensions given by the spatial and frequency sampling. In Fig.~\ref{FIG:results} our images are acquired at 64~$\times$~64 resolution, covering a field of view of 100~mm$\times$~100~mm and a frequency range of 11~GHz to 20~GHz with 201 equally spaced frequency samples (i.e. a data cube of 64~$\times$~64~$\times$~201).  In Fig.~\ref{FIG:results}(b) and (d), we image the amplitude and phase of $\hat{\delta\mathbf{r}}$ averaged over the frequency band 11-12~GHz. In Fig.~\ref{FIG:results}(c) and (e), we show the same for 19-20~GHz. 

A clear visual signature for the location of positive margins is present in Fig.~\ref{FIG:results}(b) as dark bands of $|\hat{\delta\mathbf{r}}|\sim0$ to the right and left sides of the upper and lower ramps, respectively. 
For the higher frequency band in (c), a second minimum in $|\hat{\delta\mathbf{r}}|\sim0$ occurs to right (left) of the upper (lower) ramp, i.e. for a thickness near 3~mm - this is consistent with the modelling predictions in Fig.~\ref{FIG:Setup}(c). Similarly the thickness dependence of the phase images in Fig.~\ref{FIG:results}(d) and (e) correlate well with the modelling predictions in Fig.~\ref{FIG:Setup}(b). We note that the peak values of $\hat{\delta\mathbf{r}}$ are higher than predicted for both amplitude and phase and suggest some small discrepancy between the physical and modelled system, possibly due to minor drying of the phantom between characterisation and measurement.

\begin{figure}[tp]
    \centering
\includegraphics[width=0.45\textwidth]{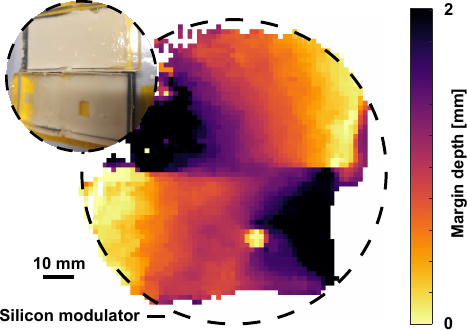} 
    \caption{Experimental demonstration of margin depth mapping. Reconstructed margin depth distribution from the full hyperspectral dataset shown in part in Fig.~\ref{FIG:results}. A photograph of the dual ramp phantom, fabricated according to the schematic in Fig.~\ref{FIG:results}(a), is shown inset. Reconstructed values yielding high error, such as those outside the phantom or photomodulator region, are omitted (white).}
    \label{FIG:Depth}
\end{figure}

To determine the margin thickness in a given location, we compare the experimental complex valued frequency spectrum $\hat{\delta\mathbf{r}}(f)$ with those predicted by the transfer matrix method for different margin depths. The margin depth is calculated by minimising the spectrally averaged distance in the complex plane between the experimental data and modelled margin depth. The full details of the algorithm are given in method section~\ref{sec:algorithm}. In Fig.~\ref{FIG:Depth}, we show the margin depth determination for the hyperspectral data set featured in Fig.~\ref{FIG:results}(b)-(e). We find the best results when we restrict the solution to the clinically relevant range of 0-2~mm, thereby circumventing false solutions due to higher order resonances and experimental artefacts (see supplementary Fig.~S7), and ensuring that margins $>$2 mm are reliably classified as negative. In Fig.~\ref{FIG:Depth}, our imaging approach is able to pick up subtle deviations from the expected ramp mould shape observed as curvature in determined margin depth. These likely appear due to compression of the phantom during fabrication. Most strikingly, however, the localised positive margin insertion is clearly identified as a bright yellow spot in 
Fig.~\ref{FIG:Depth}.

The principal result of this study is the proof of concept margin depth reconstruction in Fig.~\ref{FIG:Depth}. Within it, not only are positive margins clearly identified in yellow, but we can identify margin depths of up to 2~mm thickness. In the following sections, we evaluate the limitations of the capabilities and constraints of the approach.

\subsection{Performance and limitations}

We target an imaging resolution comparable to the histology sectioning interval of 1-3 mm. At 64$~\times~$64 pixel resolution the photoexcited pixel size is 1.5~mm~$\times~$1.5~mm, already competitive with many leading approaches to IMA, such as MRI \cite{schwarz2020,Heidkamp2021}. This coincides by design with the diffusion length of the photoexcited charge carriers in our choice of silicon photomodulator with 1.2~ms effective charge carrier lifetime. Whilst this resolution is sufficient to localise the simulated cancerous protrusion (positive margin) in Fig.~\ref{FIG:results}, higher resolution may be obtained using a lower charge carrier lifetime silicon modulator. When doing so a reduction in photomodulation efficiency is expected, but may be directly compensated for with higher optical illumination intensities.

In this feasibility study, measurement times are comparable with commercial RF probes: a 64*64 image currently takes $\sim$ 20 minutes, corresponding to 2.4 seconds per pixel data point. The evaluated subsurface volume of our phantom ($100~$mm$~\times$100~mm$\times$~2~mm) is also competitive with other approaches to IMA \cite{schwarz2020,Heidkamp2021}. There is also significant scope for reducing image acquisition times. Firstly, in our computational imaging approach the resolution, imaging area and imaging speed are interlinked parameters which may be adjusted on-the-fly to efficiently match the sample size. For example, a $32\times32$ pixel dataset acquisition would take under 5 minutes with our current system. Furthermore, we believe that measurement times can be drastically enhanced with developed systems engineering. Firstly, communication overheads with the VNA have not been optimised and currently account for $\sim$80\% of the imaging time. We are also oversampling with a higher frequency resolution than required. Taken together, a hardware (DMD) limited imaging time of 1.6 seconds for a $64\times64$ pixel image may be ultimately achievable. Finally, computational imaging is well suited to versatile approaches such as adaptive imaging \cite{Phillips2017}, integrated image processing \cite{Penketh2022} and compressive sensing \cite{Stantchev2017}, which could increase imaging speeds beyond the hardware limitation given above.

\subsection{Robustness to inter-patient variability}
It can be realistically expected that the tissue parameters we have targetted for our experimental phantoms will differ somewhat from those of fresh excised samples, as these vary throughout the literature \cite{Lazebnik2007,Kuwahara2020}.  Indeed, in supplementary Fig.~S1 we show that one can expect an improvement in imaging performance using alternative literature values for tissue permittivity. Certainly these parameters will also vary from patient to patient. We therefore consider here how robust our approach is to variations in the assumed tissue parameters using our transfer matrix method model.

\begin{figure}[h!]
    \centering
\includegraphics[width=0.50\textwidth]{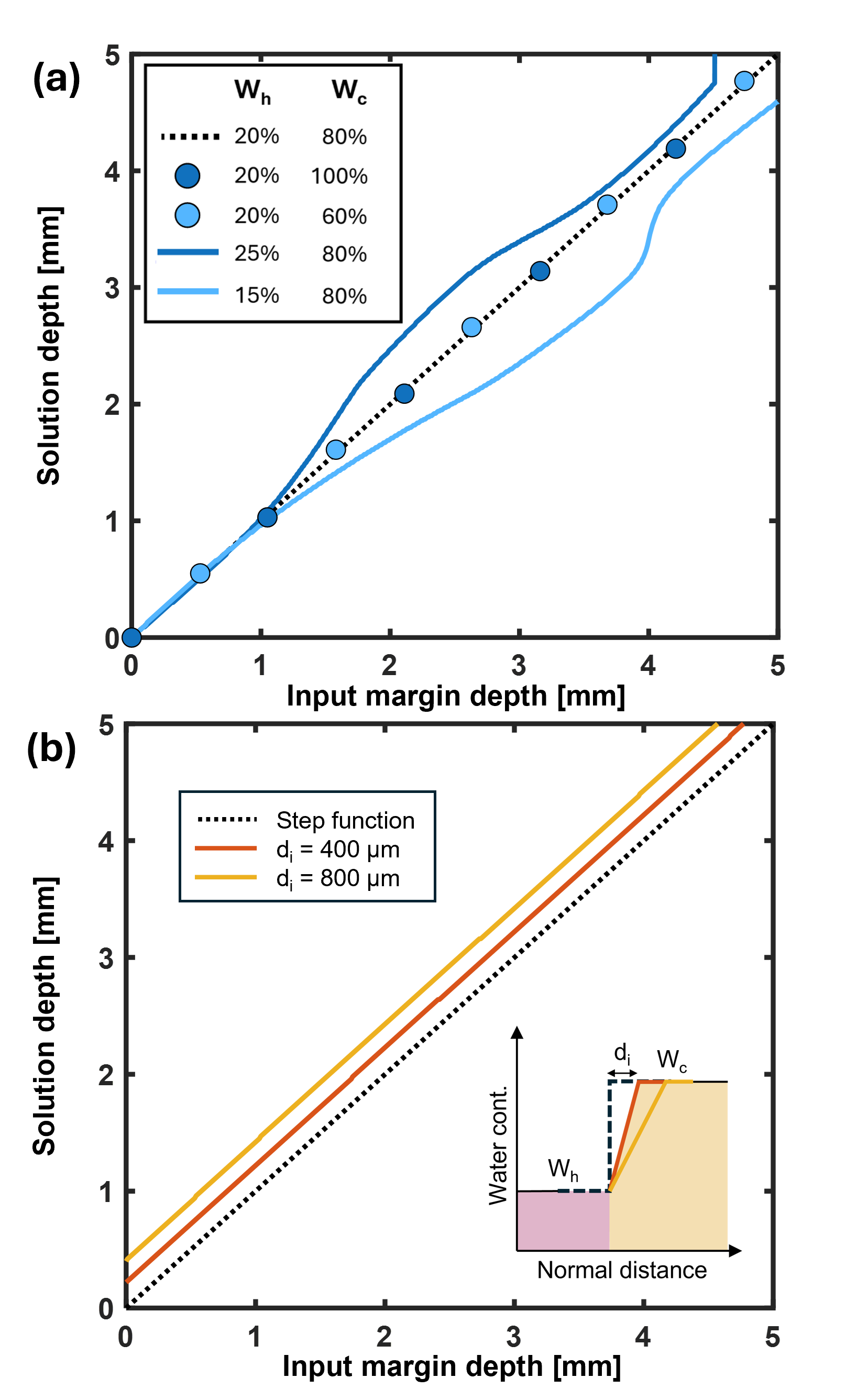} 
    \caption{
    Effect of patient-specific tissue differences on model predictions. Comparison of computed margin depths (solution depth) for simulated modified input margin $\hat{\delta\mathbf{r}}$ spectra (input margin depth). (a) Effect of input healthy ($W_h$) and cancerous ($W_c$) tissue water contents differing from those assumed in the model ($W_h = 20~\%$, $W_c = 80~\%$). The altered permittivities are obtained by linearly scaling the experimentally measured values with water content. (b) Effect of replacing the discontinuity in hydration between healthy and cancerous tissue with a linearly increasing hydration profile of width $d_i$, as shown in the inset.
    }
    \label{FIG:waterCont}
\end{figure}
In Fig.~\ref{FIG:waterCont}(a) we plot the depth of margin determined by our solver vs the ground truth simulated margin depth, for several scenarios in which the simulated margin properties differ from those assumed by the solver. This is in effect the worst case scenario, as one could incorporate fitting of the margin layer hydration levels into a more sophisticated solver. Nevertheless, when the water content of the cancerous layer is altered (circular markers), the solution is almost completely unperturbed from the ground truth dashed line solution. This almost total robustness to modest variations in the water content of the cancerous tissue (see supplementary Fig.$~$S2 and Fig.$~$S3 for the limits), suggests our approach will perform well when tested on real, varied, patient tissues. However this also means that in its current form our imaging system is not well positioned for distinguishing healthy glandular tissues from cancerous glandular tissues, as both tissues have high water contents \cite{Lazebnik2007Meas}. This is a limitation that is common to many approaches to breast screening and IMA when imaging particularly dense breasts.

When the healthy margin layer water content is varied (solid lines), the result is a slight over or under-reading of the margin thickness. However, this crucially results in only minor perturbations for the thickest of close margins of 1-2 mm, with the most clinically significant $<$1 mm depth margins unperturbed.

Similarly, it is expected that for real excised tumours the boundaries between margin and tumour tissue may be less well defined, with a gradient in hydration level between the distinct tissue regions. In  Fig.~\ref{FIG:waterCont}(b) we show that substituting a binary (step function) transition between the water contents of the margin and cancerous layers only has the effect of shifting the determined margin depth, up to the limit where the transition region width approaches 2~mm (supplementary Fig.~S5). The resulting shift is approximately half the additional separation between the original margin and cancerous regions.

\subsection{Conclusions}
We have experimentally demonstrated a novel and affordable approach to intraoperative breast cancer margin assessment, using tissue-mimicking margin phantoms. Our microwave single-pixel imaging technique provides a step change in capability over existing radiofrequency probes, offering vastly improved imaging resolution together with the ability to quantify margin depth across the excised tumour surface. Whilst there is significant scope to improve imaging speed and resolution as discussed, the demonstrated imaging timescales, resolution and detected margin depths already support clinical feasibility \cite{schwarz2020,Heidkamp2021}. In addition, our system uses continuous‑wave incoherent visible light and very low RF power ($\sim$1~mW), which is intrinsically safe and non-ionizing , while the use of an entry‑level portable vector network analyser further reduces the overall system size and hardware cost.  

 Our simulations have shown our approach to be remarkably robust to inter-patient variations in the water content of both margin and cancerous tissues, with improved imaging sensitivity expected when transitioning from phantoms to real patient tissues. By identifying inadequate margins during surgery, our imaging technique could reduce the risk of local recurrence, metastisis and the need for repeat surgeries, improving patient recovery times and survival rates, whilst reducing histological and surgical workload.

 \section{Method}
 \subsection{Breast cancer tissue phantoms}
 \label{sec:phantoms}
The tissue-mimicking hydrogel phantoms were prepared following a modified version of the recipe in \cite{Lazebnik2005}, described in detail in supplementary section 1. The 80\% water sample consists solely of water and gelatin, with the percentages expressed by weight. For the 20\% water phantom, the composition includes approximately 20\% water, 3.55\% gelatin, 73.36\% oil mixture (a 50/50 blend of safflower oil and kerosene by volume), and the remainder is surfactant.

 The permittivities of the prepared constituent phantom layers are determined by waveguide transmission / reflection measurements using a VNA. Samples from the prepared low‑water‑content and high‑water‑content phantom batches are cast into a waveguide holder and characterised between 12–18~GHz using the Nicolson–Ross–Weir method \cite{Nicolson1970,Weir1974}. The resulting permittivities can be seen in supplementary Fig.~S1(c) alongside literature values from a large scale set of clinical measurements \cite{Lazebnik2007Meas,Lazebnik2007MeasCancer} in Fig.~S1(f).

 The measured phantom layer properties differ slightly from the literature values. The $20\%$ water content margin tissue layer shown in supplementary Fig.~S1 contains a higher water content than the literature adipose tissue also shown. This makes our phantoms representative of tissue with a higher than 30$\%$ content of non-adipose, fibroglandular or fibroconnective tissue (the boundary for grouping patient data in \cite{Lazebnik2007Meas}). Greater depth penetration and imaging sensitivity is predicted for tissue with the literature values shown in supplementary Fig.~S1. Similarly the $80\%$ water content `cancerous' tissue shown in supplementary Fig.~S1 exhibits a slightly elevated water content with respect to the literature malignant breast tissue, though we note the reference values are for $>$30$\%$, not 100$\%$ malignant tissue. Additionally Fig.~\ref{FIG:waterCont}(a) shows that our technique is robust to modest hydration discrepancies for this layer. 
 \subsection{Transfer matrix modelling}
\label{sec:TMM}
The reflection from the multi-layer phantom stack is calculated using the transfer matrix method \cite{hecht2012optics}, which assumes plane wave (here at 35$^\circ$ TE) illumination and that each homogeneous layer extends infinitely in the directions parallel to the interfaces. The modelled layer properties, including a supporting substrate for the silicon wafer and plastic barrier layers to confine moisture, are described in Table~\ref{tab:TMM} below. 
\begin{table}[h!]
\centering
\caption{\bf TMM parameters}
\begin{tabular}{ccc}
\hline
Layer & Thickness [mm] & Permittivity \\
\hline
Air (incident)& 0 & 1 \\
Plas. Substr. 1& 2 & 2.25 \\
Silicon & 0.1 & 11.7 (10.0 + 96.0i)$^\dagger$ \\ 
Plas. Substr. 2& 0.1& 2.25 \\
`Margin'& $\in [0,5]$ & (7.1 + 3.9i)$^*$ \\
Plas. Barrier & 0.023 & 2.25 \\
`Cancer' & 10 & (39.7 + 27.8i)$^*$ \\
Air & 0 & 1 \\
\hline
\end{tabular}
  \label{tab:TMM}
   \caption*{Transfer matrix method model parameters. $^\dagger$ When photoexcited. $^*$ At 15 GHz - frequency dependence shown in supplementary Fig.~S1(c).}
\end{table}
$\delta \bf{r}$ is determined by subtracting the complex reflection of the photoexcited stack, with or without phantom sample as appropriate, from the non-photoexcited stack (Table~\ref{tab:TMM}). The change in permittivity of the silicon upon 100 W/m$^2$ white light photoexcitation is calculated using the Drude model following \cite{Hooper2019}, photoexcited values shown in brackets). To determine the normalised image signal ($\hat{\delta\bf{r}}$~=~$\delta\bf{r}_\text{SAM}/\delta\bf{r}_\text{REF}$), $\delta\bf{r}_\text{REF}$ is calculated in the same way including only the air, Plas. Subst. and Silicon layers in Table~\ref{tab:TMM}.

\subsection{Engineered silicon photomodulator}
 \label{sec:Si}
The photomodulator used consisted of a high resistivity $>$10,000 $\Omega$cm (100) float-zone silicon wafer that was 100 mm in diameter and 100~$\upmu$m thick. The wafer underwent a stringent wet chemical clean, which was immediately followed by transfer to the load lock of a Veeco Fiji G2 plasma-enhanced atomic layer deposition (ALD) tool. 160 cycles ($\sim$20 nm) of Al$_2$O$_3$ was deposited by ALD at 200~$\degree$C using an O$_2$ plasma source and trimethylaluminium precursor. The deposition was performed on both sides of the wafer to achieve symmetrical passivation. Following the Al$_2$O$_3$ deposition, the wafer was annealed in a quartz tube furnace for 30 min in air at 450~$\degree$C to increase the photogenerated charge carrier lifetime to $\sim$1.2~ms for the carrier densities used in this work (as shown in supplementary Fig.~S6), increasing its modulation efficiency \cite{Hooper2019}. Further details of the ALD process are described in \cite{Grant2024}.

 \subsection{Single pixel imaging}
 \label{sec:imagingMethod}
Our approach relies on the single-pixel imaging method (one that is intimately related to computational ghost imaging), in which the lack of spatial resolution provided by a single `bucket' detector is overcome by applying time-varying spatial modulations to the unknown light field \cite{Padgett2017,Gibson2020}. By combining the known sequence of projections incident on the silicon modulator (which mask the phantom) with the associated reflected microwave signal measured by the VNA, an image may be formed. Mathematically the process may be described as follows. We first vectorise the representation of the reconstructed object image $\mathrm{\mathbf{o}}$, reshaping it from a $N~\times~N$ pixels to a 1D $N^2~\times~1$ column vector. We can then write:
\begin{equation}
    \mathrm{\mathbf{o}} =  \mathrm{\mathbf{P}}^{-1} \mathrm{\mathbf{s}}.
\end{equation}
$\mathrm{\mathbf{P}}$ is a matrix representing the projected patterns (Hadamard patterns in our case). The $i^\text{th}$ column of $\mathrm{\mathbf{P}}$ holds a vectorised representation of the $i^\text{th}$ projected pattern. $\mathrm{\mathbf{s}}$ is the vector of complex-valued `bucket' signals, measuring the spatial overlap between the projections and the object or field.

In the intuitive case where $\mathrm{\mathbf{P}}$ describes the raster basis, i.e. a single `on' pixel moving with each pattern, the complex valued detector signal vector, $\mathrm{\mathbf{s}}$, represents directly an image of the photoactivated change in reflection $\delta \mathbf{r}$ at a given frequency, when reshaped to $N~\times~N$ pixels. Due to the limited photomodulation signal arising from one pixel, we opt for the more signal efficient Hadamard basis \cite{Sloane1979,harwit2012hadamard} shown in Fig.~\ref{FIG:Setup}(a).

The amplitude and phase of the electric field reflected from the modulator plus phantom stack is recorded using a portable, two port VNA (Shockline MS46122B, Anritsu). The microwave beam is tramsitted and received via a pair of waveguide-coupled lens horn antennas (18810-MB12223 UBR140, Flann Microwave), positioned with front faces $\sim$35~cm from the modulator at 35$\degree$ half angle as shown in Fig.~\ref{FIG:Setup}(a). For each of the projected patterns forming the measurement basis $\mathbf{P}$, the microwave illumination frequency is swept by the VNA over the range 11 to 20 GHz with 201 frequency points and 10~kHz intermediate frequency bandwidth. The visible photomasks are provided by a projection system utilizing a digital micromirror device (DMD) and a white xenon arc lamp  yielding an intensity of $\sim~100~$W/m$^2$.

The number of projections needed per image follows the number of pixels in the image as $2N^2$ (where the factor of two arises due to the sequential projection of positive then negative components of a single Hadamard mask, the difference in signal emulating +/- values in the masks). For the 64 $\times$ 64
pixel resolution images presented, the number of projections per hyperspectral image is 8192, taking 20 minutes to acquire. We note that a significant portion of this time is between individual spectrum acquisitions and could be reduced with synchronisation between the VNA and DMD. For the normalised results presented in Fig.~\ref{FIG:results}, no additional repeat readings were applied to the phantom sample (SAM) dataset, but 3 repeat hyperspectral images were acquired (in 1 hour) for the no object (REF) dataset.

The dominant experimental artefact in the system is the presence of spatially varying standing wave structure in the monochromatic images (see supplementary Fig.~S7), an effect which is mitigated to a degree by acquiring hyperspectral data and may be further ameliorated with radiation absorbing material, at the cost of reduced signal. As the margin depth determination algorithm necessarily returns a metric for the residual error in the `fit', we remove (white regions) portions of the FOV which are outside the dual ramp phantom by specifying a threshold error value for inclusion in the reconstruction. 

\subsection{Margin depth determination algorithm}
\label{sec:algorithm}
For each pixel of the normalised images, the total spectral error between the experiment and a range of considered margins depths (up to 2 mm) is evaluated as
\begin{equation}
\text{Err}_d = \sum_{f}|\hat{\delta \bf{r}}_{\text{expt},f}-\hat{\delta \bf{r}}_{\text{TMM},f,d}|,
\label{eq:error}
\end{equation}
i.e. distance in the complex plane, summed over frequency. The value of margin depth $d$ in the transfer matrix method modelling (TMM) which minimises `Err' is chosen as the solution and the value of `Err' can be used to threshold unacceptable solutions from the depth reconstruction. The final depth reconstructions are median filtered with $3\times3$ pixel window width to reduce high spatial frequency noise. 

Beyond 2~mm margin depth, the contrast in the resonance features diminish (Fig.~\ref{FIG:Setup}(b) and (c)) and the presence of experimental noise makes the solver increasingly susceptible to identifying the `incorrect' minimum in the error function. When the maximum solution depth is capped at 2~mm, this manifests as a small discontinuity solved thickness at $\sim1.8$~mm in Fig.~\ref{FIG:Depth} and supplementary Fig~S4. Crucially this limitation of the current experimental system arises for deep margins and does not yield false positives (see supplementary Fig.~S2 and Fig.~S3 for details). Furthermore we show in supplementary Fig.~S1 that one can expect that in real human breast tissues the imaging contrast will be nearly doubled along with the maximum determinable margin depth.

\section*{Funding}
The authors acknowledge support from the Engineering and Physical Sciences Research Council (EPSRC) via the Terabotics Programme Grant (EP/V047914/1), Computational spectral imaging in the THz band (EP/S036261/1), the TEAM-A Prosperity Partnership (EP/R004781/1), A-Meta (EP/W003341/1), the META4D Programme Grant (EP/Y015673/1) and (EP/Z535928/1). The authors also acknowledge support from the European Research Council via (804626 and 101170907).
\section*{Acknowledgements}
The authors gratefully acknowledge access to the ALD facilities provided by the Nano Fabrication Research Technology Platform at the University of Warwick. 
 
\section*{Disclosures}
The authors declare no competing financial interests.

\section*{Data Availability} 
The research data underlying this work will be made openly available upon publication of the peer‑reviewed article.
\vspace{20cm}
\newpage
\newpage
\appendix
\setcounter{figure}{0}
\renewcommand{\thefigure}{S\arabic{figure}}
\setcounter{table}{0}
\renewcommand{\thetable}{S\arabic{table}}
\section*{Supplementary Material}
This supplementary section contains supporting materials for the article ``A microwave super-resolution imaging approach towards breast cancer margin mapping''.  
\section{Tissue phantom fabrication}
\label{Sec:suppl:phantoms}
The tissue-mimicking hydrogel phantoms were prepared by modifying the recipe from \cite{Lazebnik2005} to ensure safer handling and preparation, specifically by completely removing small molecules such as p-toluic acid, n-propanol, and formaldehyde. Gelatin derived from porcine skin (Sigma-Aldrich, 48724-100G-F) was dissolved in deionized water, and a homogenized oil phase (50\% safflower oil and 50\% kerosene) was incorporated. Following the exclusion of small molecules, the ratio of each ingredient was carefully adjusted to achieve the desired hydration level while maintaining gel stability. The mixture was vigorously stirred, and a surfactant was added to ensure emulsion stability. After cooling to approximately 34 °C, the mixture was poured into moulds and allowed to solidify as the gelatin matrix congealed overnight. This modification preserves the capability to fabricate heterogeneous phantoms that remain stable over extended periods, exhibiting no discernible solute or solvent diffusion even when materials with differing water content are placed in direct contact, as demonstrated within \cite{Lazebnik2005}. Furthermore, the recipe is highly suitable for broadband microwave applications, as the dielectric properties can be precisely tuned by adjusting the oil content in the gelatin matrix, enabling accurate simulation of various biological tissues for advanced imaging studies. Unlike the original protocol, which relied on chemical cross-linkers for long-term stability, we observed that vacuum sealing the phantoms effectively preserves hydration, prevents mould growth, and maintains both mechanical and dielectric properties, supporting reliable performance in quantitative microwave imaging experiments.

The 80$\%$ water sample consists solely of water and gelatin, with the percentages expressed by weight. For the 20$\%$ water phantom, the composition includes approximately 20$\%$ water, 3.55$\%$ gelatin, 73.36$\%$ oil mixture (a 50/50 blend of safflower oil and kerosene by volume), and the remainder is surfactant (Fairy liquid), all percentages given by weight. A critical step in preparing the phantom samples to ensure they are free of air bubbles involves gently folding the mixture manually in a figure-eight motion within the beaker after the surfactant has been incorporated. This careful manual mixing promotes uniformity and minimizes entrapped air, which is essential for the phantom’s optical and dielectric consistency.

\section{Tissue phantom optical parameters}
\begin{figure*}[]
    \centering
\includegraphics[width=1\textwidth]{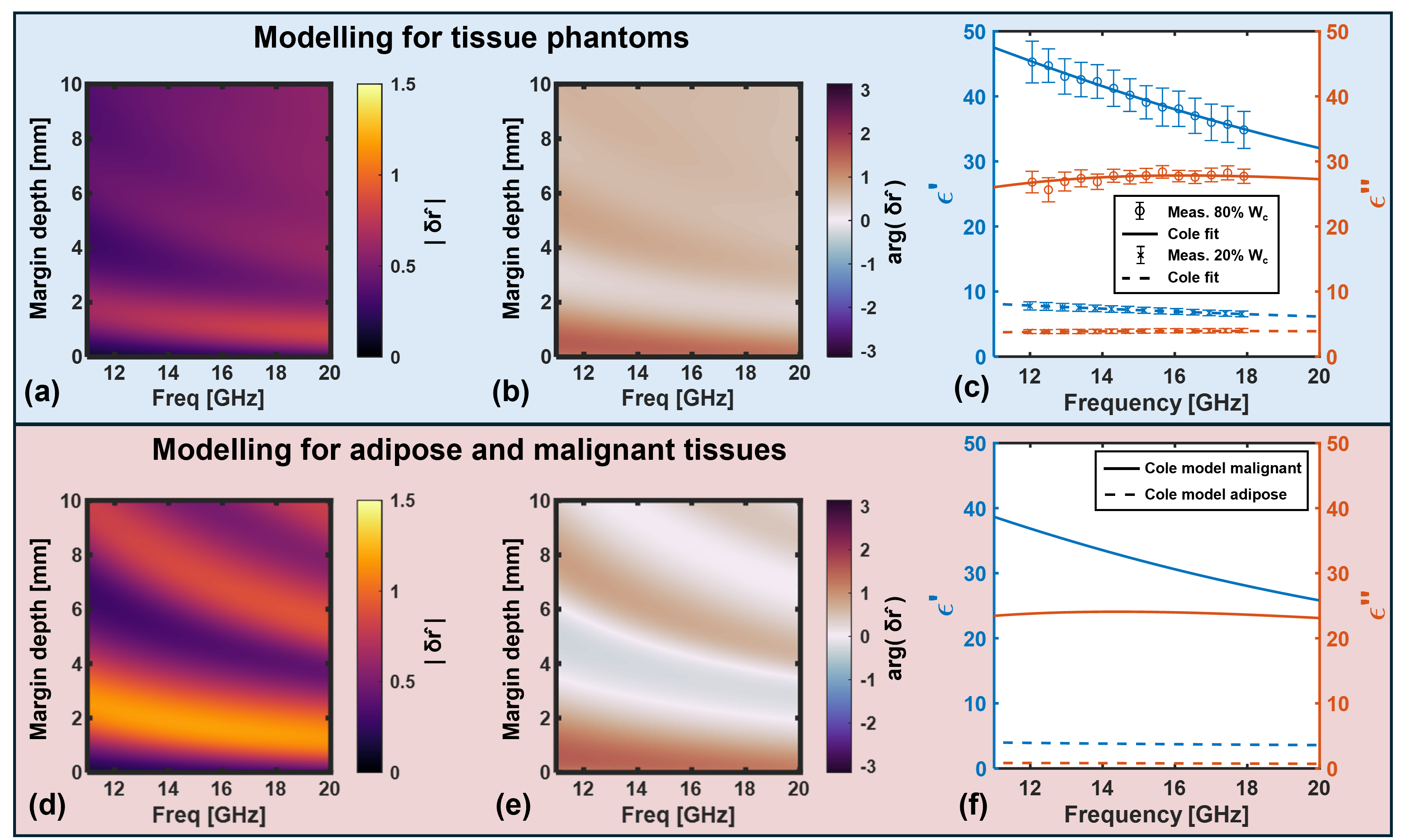} 
    \caption{A comparison of the modelled imaging system sensitivity for the fabricated phantoms used in this work with literature cole-cole parameters for breast tissues \cite{Lazebnik2007MeasCancer}. (a) and (b) show the normalised phase and amplitude of the complex reflection coefficient predicted from transfer matrix modelling, for layered phantoms as described in Table~1 of the main text and with complex layer permittivity ($\epsilon=\epsilon'$+$i\epsilon''$) given by the cole-cole model fits in (c). In (d) - (f) the phantom layers permittivities from \cite{Lazebnik2007MeasCancer} are cole-cole fits to measured human breast that contain 30$\%$ or more malignant tissue (malignant) or 70$\%$ or more adipose tissue (adipose).}
    \label{SFIG:tissueCom}
\end{figure*}
\begin{table*}[ht]
\centering
\caption{Fitted electrical parameters for phantom and literature tissue models.}
\label{tab:material_params}
\begin{tabular}{lccccc}
\hline
Material &
$\epsilon_{\infty}$  &
$\Delta \epsilon$  &
$\tau$ (s) &
$\alpha$ &
$\sigma_s$ (S/m) \\
\hline
Cancerous phantom
& 9.7689
& 53.9989
& $9.4979\times10^{-12}$
& 0.0001
& 0.77215 \\

Healthy phantom
& 2.2640
& 8.3326
& $8.6045\times10^{-12}$
& 0.0763
& 0.23426 \\

Healthy (literature)
& 3.140
& 1.708
& $1.465\times10^{-11}$
& 0.061
& 0.036 \\

Cancerous (literature)
& 6.749
& 50.09
& $1.050\times10^{-11}$
& 0.051
& 0.794 \\
\hline
\end{tabular}
\end{table*}

The measured tissue phantoms, for the 20$\%$ water content healthy and 80$\%$ water content cancerous tissue phantoms can be seen in Fig.~S1 alongside the corresponding modelling predictions for $\hat{\delta\textbf{r}}$. The experimental complex relative permittivity is fitted using the Cole--Cole formulation with an
explicit conductivity term,
\begin{equation}
\varepsilon_r(\omega)
=
\varepsilon_{\infty}
+
\frac{\Delta\varepsilon}{1+\left(i\omega\tau\right)^{1-\alpha}}
-
\frac{i\,\sigma_s}{\omega\varepsilon_0},
\label{eq:cole-cole}
\end{equation}
where $\varepsilon_{\infty}$ is the high-frequency permittivity,
$\Delta\varepsilon$ is the dielectric dispersion strength,
$\tau$ is the relaxation time,
$\alpha$ is the Cole--Cole broadening parameter,
$\sigma_s$ is the static conductivity,
$\varepsilon_0$ is the permittivity of free space,
and $\omega = 2\pi f$ is the angular frequency. The fitted values may be seen in supplementary Table S1.

In Fig.~S1(f) we show commonly accept literature values for fatty and malignant tissue, and accordingly that the modelled imaging sensitivity would be increased for typical human tissues.

\begin{figure*}[htbp!]
    \centering
\includegraphics[width=0.85\textwidth]{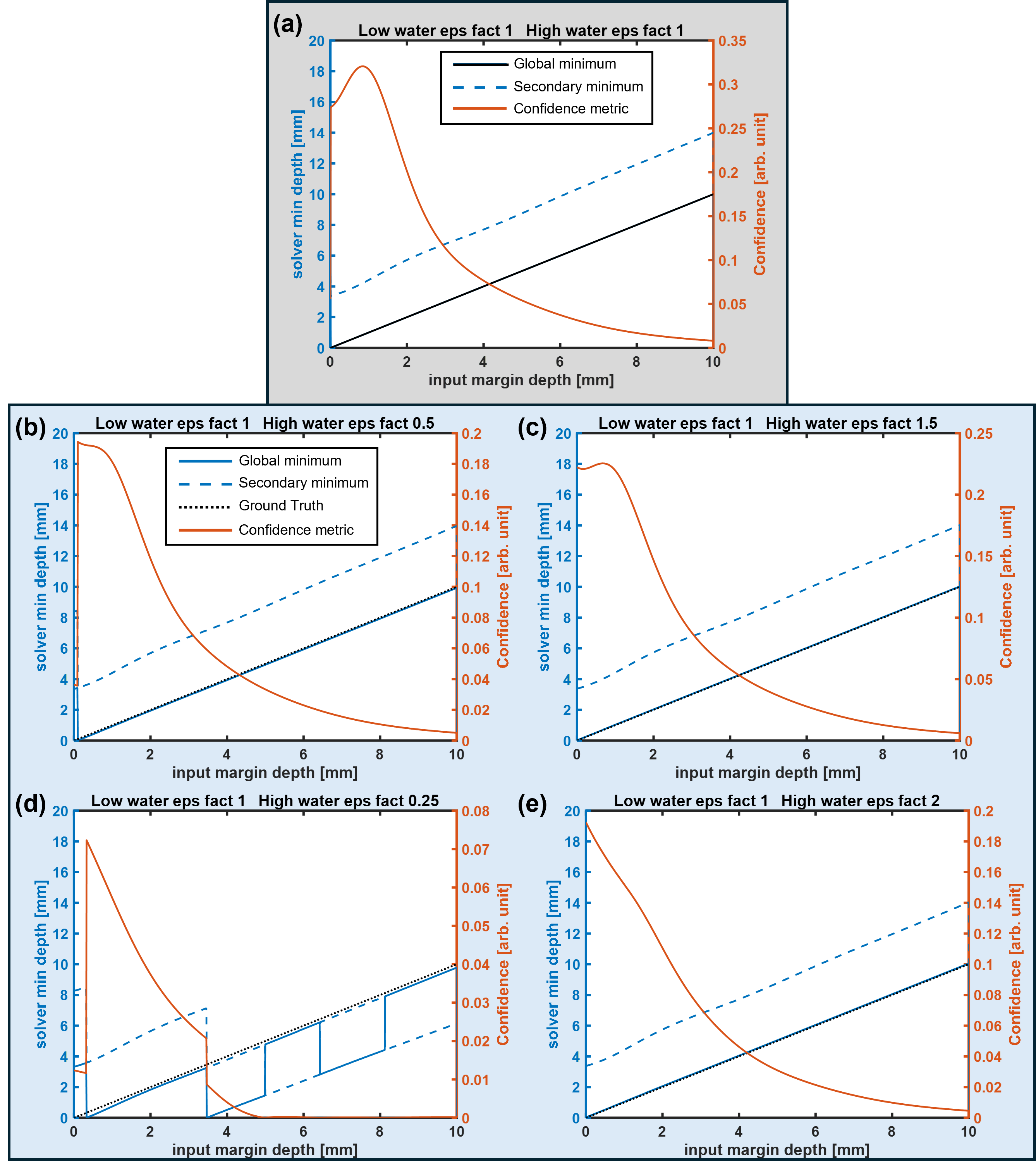} 
    \caption{Expanded effect of patient specific variations in cancerous tissue hydration level on the reconstructed margin depth. Comparison of computed margin depths (solution depth), for simulated modified input margin $\hat{\delta\mathbf{r}}$ spectra (input margin depth) with differing permittivities. (a) where cancerous tissue water content W$_c$ matches that assumed in the model. In (b)-(e) the modified permittivities are given by a linear scaling (High\_water\_eps\_fact) of the experimentally measured values by the factor indicated in each subplot title. Note therefore that this is an approximation not intended to accurately mimic real tissues. The solution with the second lowest minimum in the cost function (Err - eq. 2 in the main text) is also shown in each case. A confidence metric is also plotted which is equal to the difference in cost function between the primary and secondary minima.
    }
    \label{SFIG:HighWaterSweep}
\end{figure*}
\begin{figure*}[htbp!]
    \centering
\includegraphics[width=0.85\textwidth]{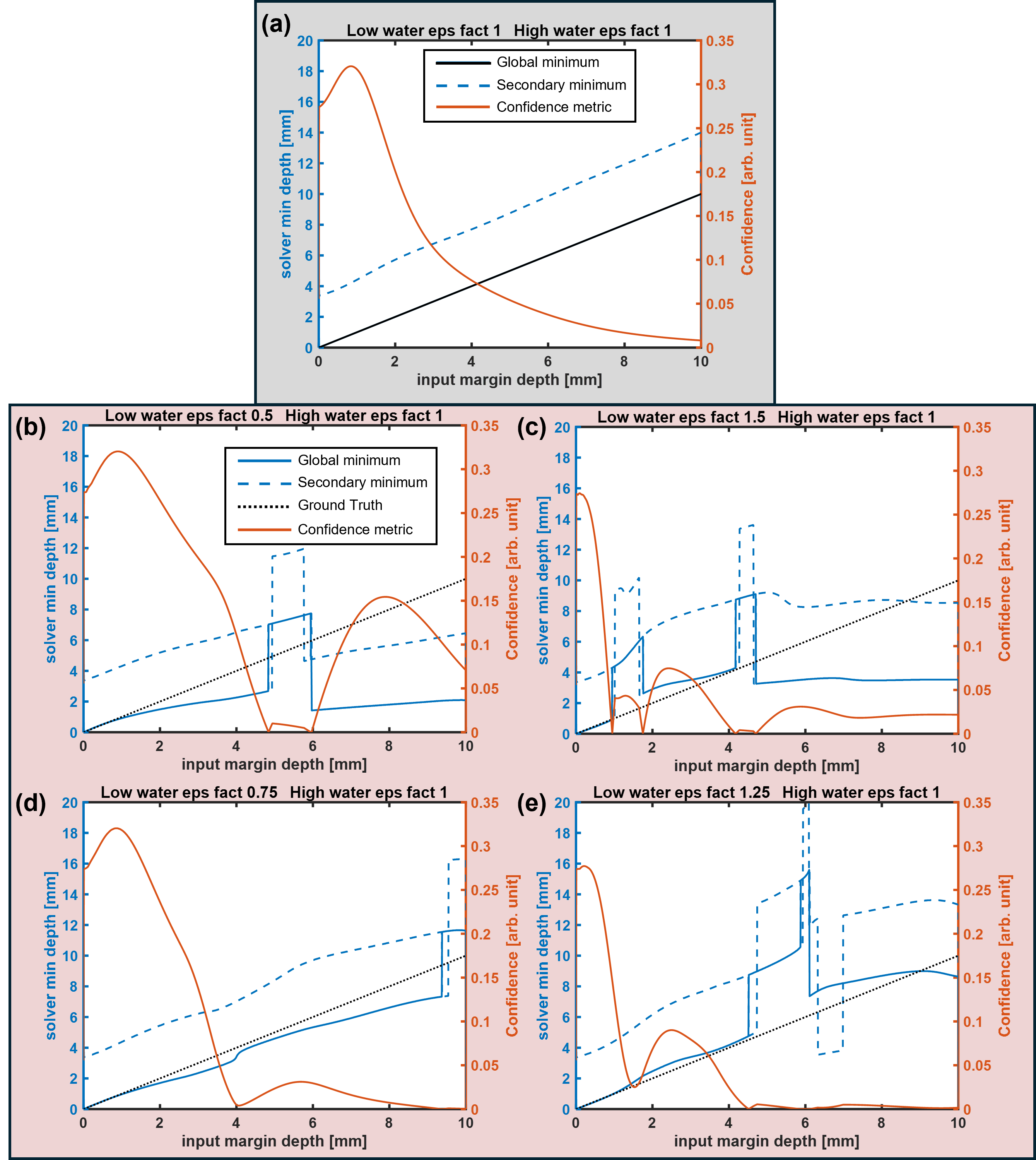} 
    \caption{As in Fig.~S2 but for variations in healthy (margin) tissue hydration level on the reconstructed margin depth. (a) where healthy tissue water content W$_h$ matches that assumed in the model. In (b)-(e) the modified permittivities are given by a linear scaling (Low\_water\_eps\_fact) of the experimentally measured values by the factor indicated in each subplot title. 
    }
    \label{SFIG:LowWaterSweep}
\end{figure*}
\begin{figure*}[htbp!]
    \centering
\includegraphics[width=0.65\textwidth]{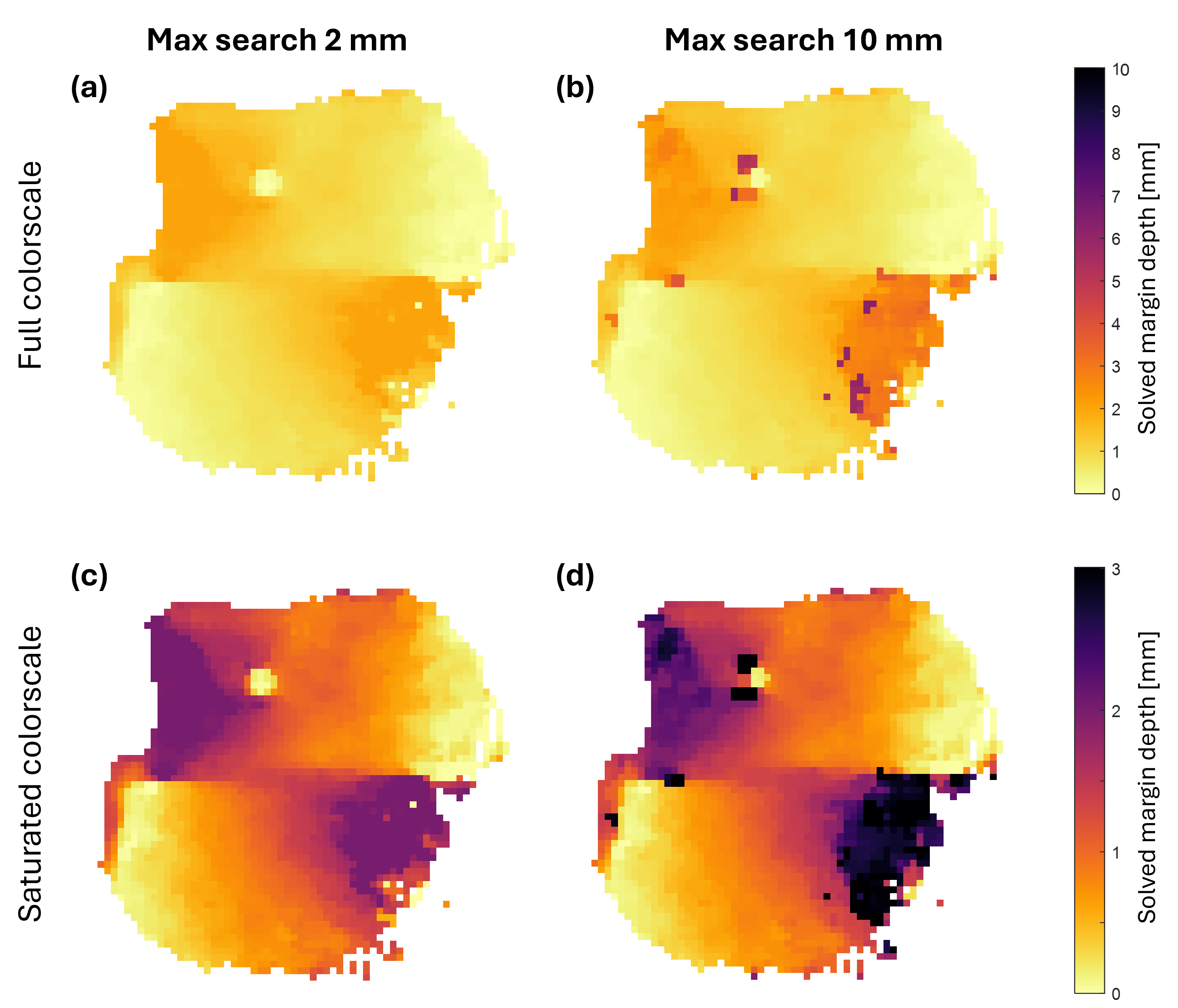} 
    \caption{Restricting the maximum margin search depth leads to more robust clinically relevant information. When essentially uncapped at 10 mm maximum margin depth, in (b) and (d), deep margins $\sim$2-3~mm and pixels spanning abrupt changes in margin depth may overestimate the margin depth. When the algorithm solutions are restricted to 0-2~mm, in (a) and (c), the distinction between negative margins can be distinguished reliably. In (c) and (d) the margin depth colourscale is saturated for enhanced visual contrast. 
    }
    \label{SFIG:solverCap}
\end{figure*}

\begin{figure}[htbp!]
    \centering
\includegraphics[width=0.45\textwidth]{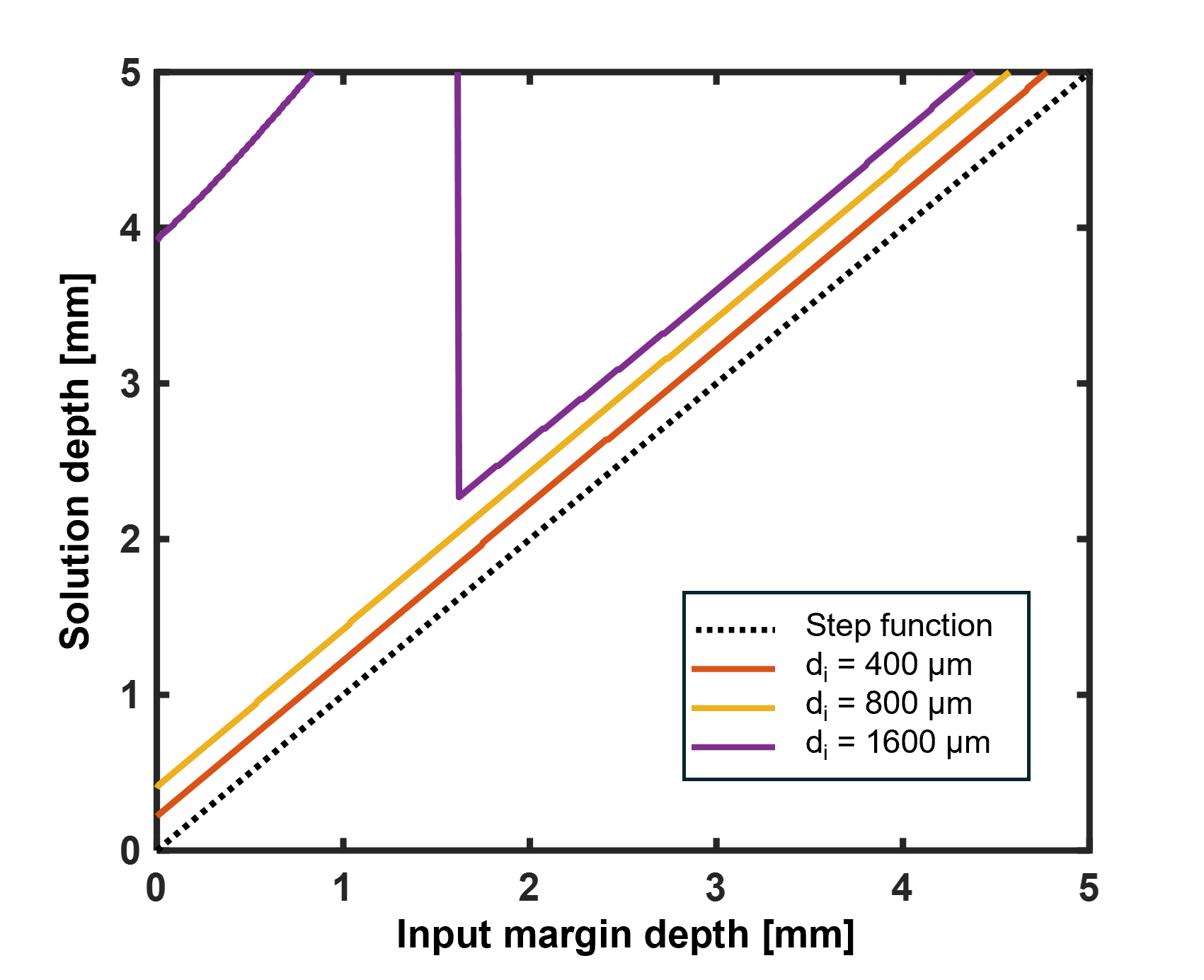} 
    \caption{As in main text Fig.~4(b) showing that substituting a binary (step function) transition between the water contents of the margin and cancerous layers only has the effect of shifting the determined margin depth, up until the transition width approaches 2~mm and the solver fails dramatically.}
    \label{SFIG:borderWidth}
\end{figure}
\section{Margin depth solver}
In Fig.~4 of the main text, the range over which the assumed permittivities differ from the solver input is restricted to $\pm$25$\%$ for visual clarity. In Fig.~S2 and S3, we show an expanded form of the data which includes the eventual failure of the solver. In addition we gain insights into the failure mode by plotting the second best solution (secondary minimum in solver cost function) found by the solver, which eventually coincides with and then replaces the correct solution. One interesting observation here is that the confidence in the solved margin depth, determined by the difference in cost function depth with the next best solution, generally remains very high for margins less than 1~mm, which are the most likely to require reoperations.

In Fig.~S5 a similar extension is made to the width over which the input margin transition zone is increased, leading to dramatic failure once the boundary width reaches 1.6~mm. Finally, Fig.~S4 shows that restricting the allowed solutions of the margin depth solver to the clinical target minimum margin depth of 2~mm produces are more robust margin map, by eliminating potential ambiguities with higher order modes (see main text Fig.~1(b) and (c)).

\section{Photomodulator characterisation}
Fig.~S6 shows the effective charge carrier lifetime of photogenerated carriers in the wafer. The effective carrier lifetime was measured by the photoconductance decay method using Sinton Instruments WCT-120PL lifetime system in transient mode, averaged over 5 successive measurements. In this work an illumination intensity of 100~W/m$^2$ yields an effective carrier lifetime near the low injection level plateau of $\approx$1.2~ms as shown in Fig.~S6. 
\begin{figure}[htbp!]
    \centering
\includegraphics[width=0.50\textwidth]{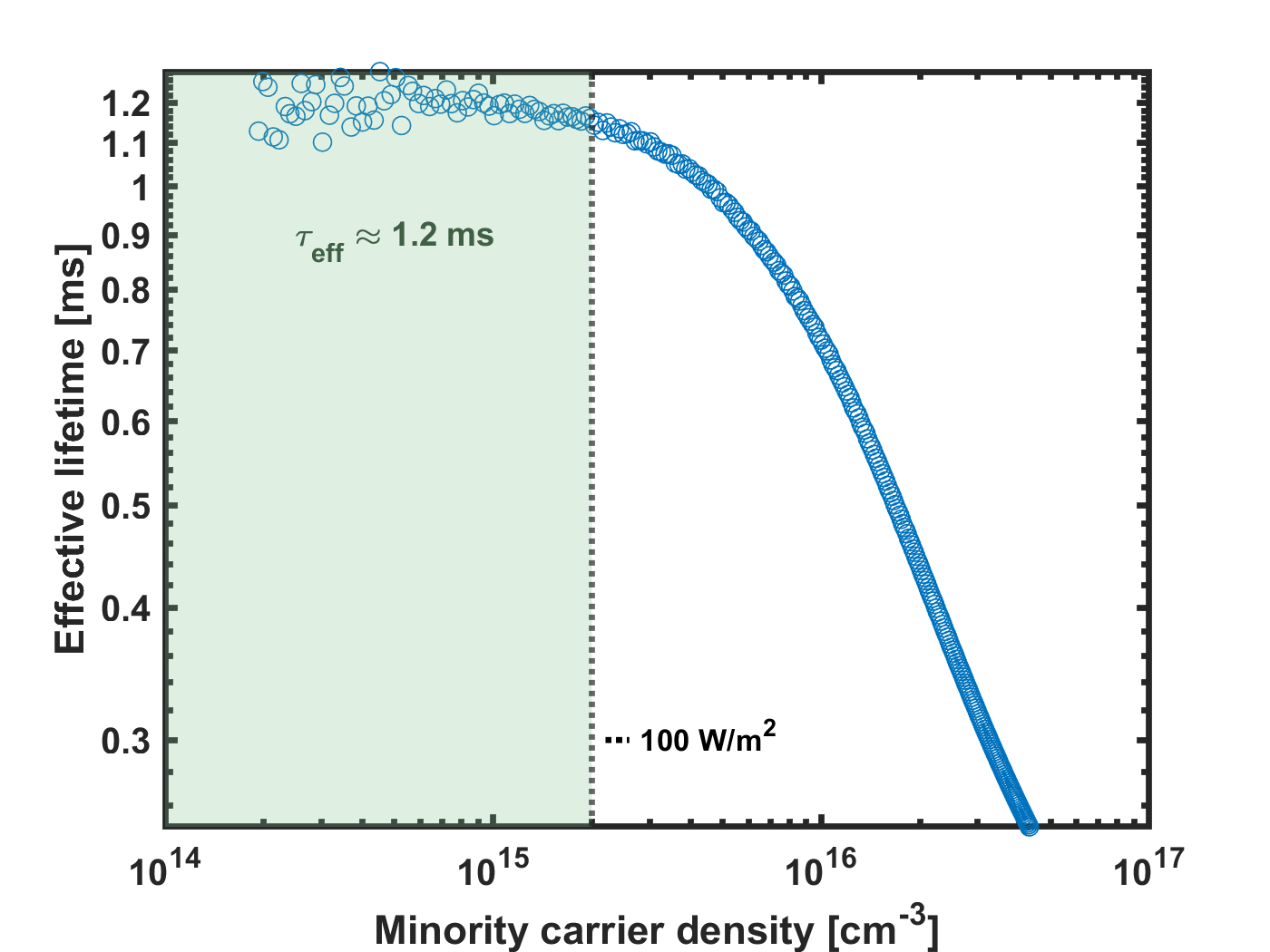} 
    \caption{The effective carrier lifetime of the silicon photomodulator used in this work, as measured by the photoconductance decay method. The carrier injection level at an illumination intensity of 100~W/m$^2$ as used for imaging is indicated by the dashed black line. The low injection level plateau where $\tau_{\text{eff}}$$\approx$1.2~ms is shaded in green.}
    \label{SFIG:waferLife}
\end{figure}

\begin{figure*}[htbp!]
    \centering
\includegraphics[width=0.97\textwidth]{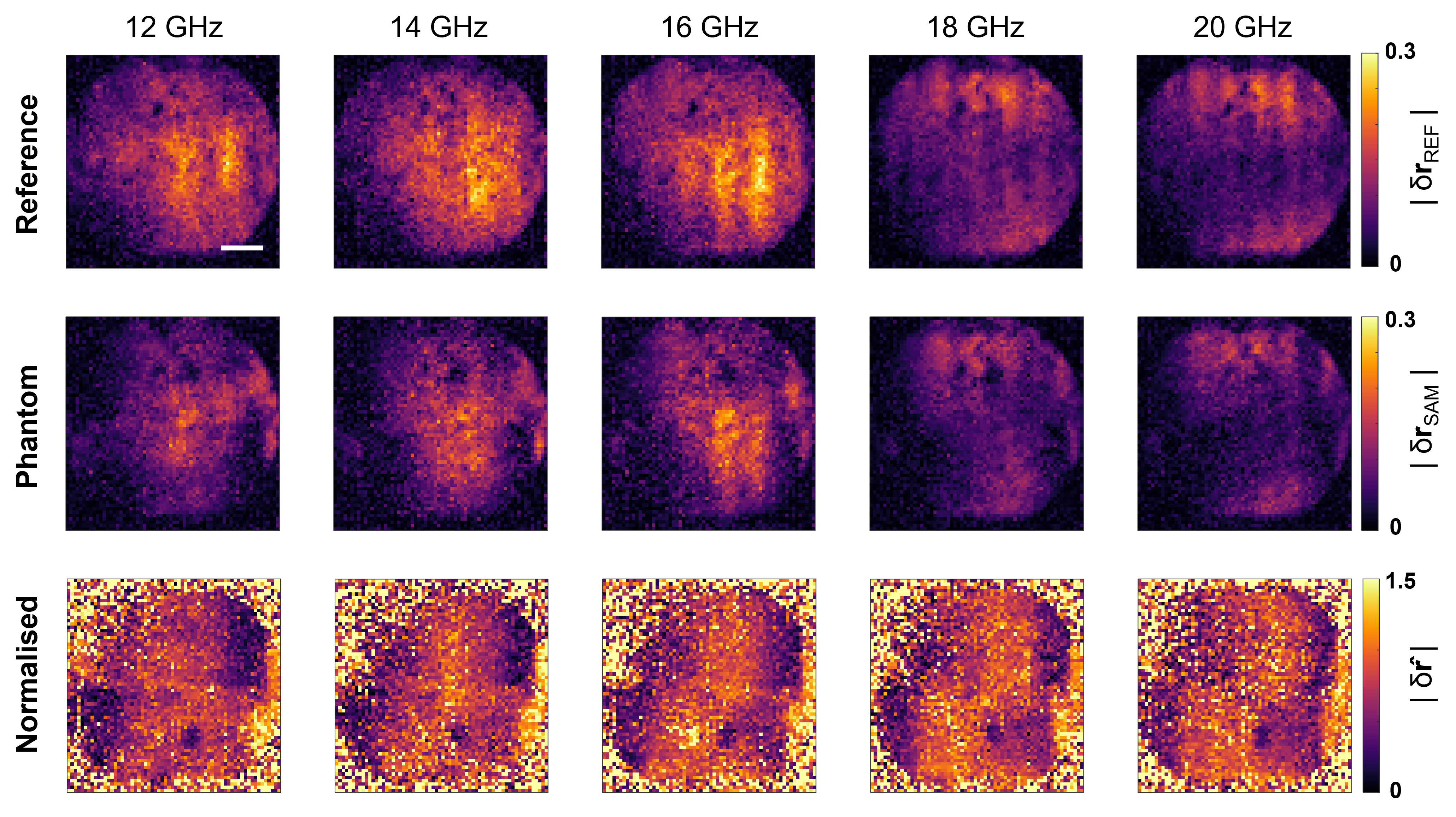} 
    \caption{Example raw image data forming part of the hyperspectral dataset. The columns indicate the single frequency at which the image was acquired (in contrast to main text Fig~2(b) and (c) which are spectrally averaged). In the top row are the reference images without the phantom sample, quantifying spatial inhomogeneities in the visible and microwave illumination and silicon wafer modulation depth. Vertical stripes artefacts are visible due to the presence of standing waves in the imaging system. In the middle row the same images are taken with the dual ramp phantom present. When normalised in the bottom row, the phantom structure is revealed. The white scale bar corresponds to  2~cm.
    }
    \label{SFIG:rawImages}
\end{figure*}
\section{Image normalisation}
Finally, for completeness, in Fig.~S7 we show the unnormalised $|\hat{\delta \bf{r}}|$ images without averaging repeat measurements, spectrally averaging, or spatially median filtering. In this case the presence of experimental artefacts such as standing waves (vertical bands) may be observed. 

\bibliography{sample}

@article{Chung_2008,
doi = {10.1088/0031-9155/53/23/005},
url = {https://dx.doi.org/10.1088/0031-9155/53/23/005},
year = {2008},
month = {nov},
publisher = {},
volume = {53},
number = {23},
pages = {6713},
author = {Chung, S H and Cerussi, A E and Klifa, C and Baek, H M and Birgul, O and Gulsen, G and Merritt, S I and Hsiang, D and Tromberg, B J},
title = {In vivo water state measurements in breast cancer using broadband diffuse optical spectroscopy},
journal = {Physics in Medicine \& Biology}
}

@article{Hubbard2021,
author = {Hubbard, Thomas J. E. and Dudgeon, Alexander P. and Ferguson, Douglas J. and Shore, Angela C. and Stone, Nicholas},
title = {Utilization of Raman spectroscopy to identify breast cancer from the water content in surgical samples containing blue dye},
journal = {Translational Biophotonics},
volume = {3},
number = {2},
pages = {e202000023},
doi = {https://doi.org/10.1002/tbio.202000023},
url = {https://onlinelibrary.wiley.com/doi/abs/10.1002/tbio.202000023},
year = {2021}
}

@article{Hussein2019,
   author = {Mousa Hussein and Falah Awwad and Dwija Jithin and Husain El Hasasna and Khawlah Athamneh and Rabah Iratni},
   doi = {10.1038/s41598-019-41124-1},
   issn = {2045-2322},
   issue = {1},
   journal = {Scientific Reports},
   month = {3},
   pages = {4681},
   title = {Breast cancer cells exhibits specific dielectric signature in vitro using the open-ended coaxial probe technique from 200~{MH}z to 13.6~{GH}z},
   volume = {9},
   year = {2019},
}

@article{Tang2017,
   author = {Sarah Shuk-Kay Tang and Sarantos Kaptanis and James B. Haddow and Giuseppina Mondani and Beatrix Elsberger and Marios Konstantinos Tasoulis and Christine Obondo and Neil Johns and Wisam Ismail and Asim Syed and Panayioti Kissias and Mary Venn and Souganthy Sundaramoorthy and Gareth Irwin and Amtul S. Sami and Dalia Elfadl and Alice Baggaley and Dionysios Dennis Remoundos and Fiona Langlands and Petros Charalampoudis and Zoe Barber and Werbena L.S. Hamilton-Burke and Ayesha Khan and Chiara Sirianni and Louise Anne-Marie Grant Merker and Sunita Saha and Risha Arun Lane and Sharat Chopra and Sophie Dupré and Aidan T. Manning and Edward R. St John and Aya Musbahi and Nokwanda Dlamini and Caitlin L. McArdle and Chloe Wright and James O. Murphy and Ravi Aggarwal and Matei Dordea and Karen Bosch and Donna Egbeare and Hisham Osman and Salim Tayeh and Faraz Razi and Javeria Iqbal and Serena F.C. Ledwidge and Vanessa Albert and Yazan Masannat},
   doi = {10.1016/j.ejca.2017.07.032},
   issn = {09598049},
   journal = {European Journal of Cancer},
   month = {10},
   pages = {315-324},
   title = {Current margin practice and effect on re-excision rates following the publication of the SSO-ASTRO consensus and ABS consensus guidelines: a national prospective study of 2858 women undergoing breast-conserving therapy in the UK and Ireland},
   volume = {84},
   year = {2017},
}

@techreport{ABS2024,
  author = {{Association of Breast Surgery}},
  title = {Recommendations for the management of radial surgical margins in patients undergoing breast-conserving surgery for ductal carcinoma in situ (DCIS)},
  year = {2024},
  url = {https://associationofbreastsurgery.org.uk/media/3z5bk0nj/dcis-margins-guidance-2024-v2.pdf},
  note = {Accessed: 1 May 2025}
}

@article{Pilewskie2018,
   author = {Melissa Pilewskie and Monica Morrow},
   doi = {10.1002/cncr.31221},
   issn = {10970142},
   issue = {7},
   journal = {Cancer},
   month = {4},
   pages = {1335-1341},
   pmid = {29338088},
   publisher = {John Wiley and Sons Inc.},
   title = {Margins in breast cancer: How much is enough?},
   volume = {124},
   year = {2018},
}

@article{Martellosio2015,
   author = {A. Martellosio and M. Pasian and M. Bozzi and L. Perregrini and A. Mazzanti and F. Svelto and P. E. Summers and G. Renne and M. Bellomi},
   doi = {10.1049/el.2015.1199},
   issn = {00135194},
   issue = {13},
   journal = {Electronics Letters},
   month = {6},
   pages = {974-975},
   publisher = {Institution of Engineering and Technology},
   title = {0.5-50 GHz dielectric characterisation of breast cancer tissues},
   volume = {51},
   year = {2015},
}

@article{Heidkamp2021,
   author = {Jan Heidkamp and Mirre Scholte and Camiel Rosman and Srirang Manohar and Jurgen J. Fütterer and Maroeska M. Rovers},
   doi = {10.1002/ijc.33570},
   issn = {10970215},
   issue = {3},
   journal = {International Journal of Cancer},
   month = {8},
   pages = {635-645},
   pmid = {33739453},
   publisher = {John Wiley and Sons Inc},
   title = {Novel imaging techniques for intraoperative margin assessment in surgical oncology: A systematic review},
   volume = {149},
   year = {2021},
}

@article{Schwarz2020,
   author = {Julia Schwarz and Hank Schmidt},
   doi = {10.1245/s10434-020-08483-w},
   issn = {15344681},
   issue = {7},
   journal = {Annals of Surgical Oncology},
   month = {7},
   pages = {2278-2287},
   pmid = {32350717},
   publisher = {Springer},
   title = {Technology for Intraoperative Margin Assessment in Breast Cancer},
   volume = {27},
   year = {2020},
}

@article{Hooper2019,
   author = {I. R. Hooper and N. E. Grant and L. E. Barr and S. M. Hornett and J. D. Murphy and E. Hendry},
   doi = {10.1038/s41598-019-54011-6},
   issn = {20452322},
   issue = {1},
   journal = {Scientific Reports},
   month = {12},
   pmid = {31797937},
   publisher = {Nature Research},
   title = {High efficiency photomodulators for millimeter wave and THz radiation},
   volume = {9},
   year = {2019},
}

@article{Phillips2017,
author = {David B. Phillips  and Ming-Jie Sun  and Jonathan M. Taylor  and Matthew P. Edgar  and Stephen M. Barnett  and Graham M. Gibson  and Miles J. Padgett },
title = {Adaptive foveated single-pixel imaging with dynamic supersampling},
journal = {Science Advances},
volume = {3},
number = {4},
pages = {e1601782},
year = {2017},
doi = {10.1126/sciadv.1601782},
URL = {https://www.science.org/doi/abs/10.1126/sciadv.1601782}}

@article{Stantchev2017,
author = {Rayko I. Stantchev and David B. Phillips and Peter Hobson and Samuel M. Hornett and Miles J. Padgett and Euan Hendry},
journal = {Optica},
number = {8},
pages = {989--992},
publisher = {Optica Publishing Group},
title = {Compressed sensing with near-field THz radiation},
volume = {4},
month = {Aug},
year = {2017},
url = {https://opg.optica.org/optica/abstract.cfm?URI=optica-4-8-989},
doi = {10.1364/OPTICA.4.000989},
}

@article{Penketh2022,
  title={Implicit image processing with ghost imaging},
  author={Penketh, Harry and Barnes, William L and Bertolotti, Jacopo},
  journal={Optics Express},
  volume={30},
  number={5},
  pages={7035--7043},
  year={2022},
  publisher={Optica Publishing Group}
}

@article{Padgett2017,
   author = {Miles J. Padgett and Robert W. Boyd},
   doi = {10.1098/rsta.2016.0233},
   issn = {1364503X},
   issue = {2099},
   journal = {Philosophical Transactions of the Royal Society A: Mathematical, Physical and Engineering Sciences},
   month = {8},
   pmid = {28652490},
   publisher = {Royal Society Publishing},
   title = {An introduction to ghost imaging: Quantum and classical},
   volume = {375},
   year = {2017},
}

@article{Gibson2020,
   author = {Graham M. Gibson and Steven D. Johnson and Miles J. Padgett},
   doi = {10.1364/oe.403195},
   issn = {10944087},
   issue = {19},
   journal = {Optics Express},
   month = {9},
   pages = {28190},
   pmid = {32988095},
   publisher = {Optica Publishing Group},
   title = {Single-pixel imaging 12 years on: a review},
   volume = {28},
   year = {2020},
}

@article{Houssami2014Margins,
  author    = {Nehmat Houssami and Petra Macaskill and M. Luke Marinovich and Monica Morrow},
  title     = {The Association of Surgical Margins and Local Recurrence in Women with Early-Stage Invasive Breast Cancer Treated with Breast-Conserving Therapy: A Meta-Analysis},
  journal   = {Annals of Surgical Oncology},
  volume    = {21},
  number    = {3},
  pages     = {717--730},
  year      = {2014},
  doi       = {10.1245/s10434-014-3480-5},
  publisher = {Springer}
}

@article{Ha2018,
  author = {Ha, R and Friedlander, L C and Hibshoosh, H and others},
  title = {Optical coherence tomography: a novel imaging method for post-lumpectomy breast margin assessment - a multi-reader study},
  journal = {Academic Radiology},
  year = {2018},
  volume = {25},
  number = {3},
  pages = {279--287}
}

@article{Allen2018,
  author = {Allen, W M and Foo, K Y and Zilkens, R and others},
  title = {Clinical feasibility of optical coherence micro-elastography for imaging tumor margins in breast-conserving surgery},
  journal = {Biomedical Optics Express},
  year = {2018},
  volume = {9},
  number = {12},
  pages = {6331--6349}
}

@article{Assayag2014,
  author = {Assayag, O and Antoine, M and Sigal-Zafrani, B and others},
  title = {Large field, high resolution full-field optical coherence tomography: a pre-clinical study of human breast tissue and cancer assessment},
  journal = {Technology in Cancer Research \& Treatment},
  year = {2014},
  volume = {13},
  number = {5},
  pages = {455--468}
}

@article{Chin2017,
  author = {Chin, L and Latham, B and Saunders, C M and Sampson, D D and Kennedy, B F},
  title = {Simplifying the assessment of human breast cancer by mapping a micro-scale heterogeneity index in optical coherence elastography},
  journal = {Journal of Biophotonics},
  year = {2017},
  volume = {10},
  number = {5},
  pages = {690--700}
}

@article{Brachtel2016,
  author = {Brachtel, EF and Johnson, NB and Huck, AE and others},
  title = {Spectrally encoded confocal microscopy for diagnosing breast cancer in excision and margin specimens},
  journal = {Lab Investig.},
  year = {2016},
  volume = {96},
  number = {4},
  pages = {459--467}
}

@article{Cahill2018,
  author = {Cahill, LC and Giacomelli, MG and Yoshitake, T and others},
  title = {Rapid virtual hematoxylin and eosin histology of breast tissue specimens using a compact fluorescence nonlinear microscope},
  journal = {Lab Investig.},
  year = {2018},
  volume = {98},
  number = {1},
  pages = {150--160}
}

@article{ChangTP,
  author = {Chang, TP and Leff, DR and Shousha, S and others},
  title = {Imaging breast cancer morphology using probe-based confocal laser endomicroscopy: towards a real-time intraoperative imaging tool for cavity scanning},
  doi = {10.1007/s10549-015-3543-8},
  journal = {Breast Cancer Res Treat},
  year = {2015}
}

@article{Tao2014,
  author = {Tao, YK and Shen, D and Sheikine, Y and others},
  title = {Assessment of breast pathologies using nonlinear microscopy},
  journal = {Proc Natl Acad Sci U S A},
  year = {2014},
  volume = {111},
  number = {43},
  pages = {15304--15309}
}

@article{Vyas2017,
  author = {Vyas, K and Hughes, M and Leff, DR and Yang, GZ},
  title = {Methylene-blue aided rapid confocal laser endomicroscopy of breast cancer},
  journal = {J Biomed Opt.},
  year = {2017},
  volume = {22},
  number = {2},
  pages = {020501}
}

@article{Yoshitake2016,
  author = {Yoshitake, T and Giacomelli, MG and Cahill, LC and others},
  title = {Direct comparison between confocal and multiphoton microscopy for rapid histopathological evaluation of unfixed human breast tissue},
  journal = {J Biomed Opt.},
  year = {2016},
  volume = {21},
  number = {12},
  pages = {126021}
}

@article{Wang2017Raman,
  author = {Wang, Y and Reder, N P and Kang, S and others},
  title = {Raman-encoded molecular imaging with topically applied SERS nanoparticles for intraoperative guidance of lumpectomy},
  journal = {Cancer Research},
  year = {2017},
  volume = {77},
  number = {16},
  pages = {4506--4516}
}

@article{Stone2021,
  author = {Stone, N and Matousek, P and others},
  title = {Spatially offset Raman spectroscopy},
  journal = {Nature Reviews Methods Primers},
  year = {2021}
}

@article{Haskell2023,
  author       = {Jennifer Haskell and Thomas Hubbard and Claire Murray and Benjamin Gardner and Charlotte Ives and Douglas Ferguson and Nick Stone},
  title        = {High wavenumber Raman spectroscopy for intraoperative assessment of breast tumour margins},
  journal      = {The Analyst},
  year         = {2023},
  volume       = {148},
  number       = {18},
  pages        = {3974--3983},
  doi          = {10.1039/D3AN00574G},
  publisher    = {Royal Society of Chemistry},
  url          = {https://doi.org/10.1039/D3AN00574G}
}

@article{McClatchy2018,
  author       = {McClatchy 3rd, D. M. and Zuurbier, R. A. and Wells, W. A. and Paulsen, K. D. and Pogue, B. W.},
  title        = {Micro-computed tomography enables rapid surgical margin assessment during breast conserving surgery (BCS): correlation of whole BCS micro-CT readings to final histopathology},
  journal      = {Breast Cancer Research and Treatment},
  year         = {2018},
  volume       = {172},
  number       = {3},
  pages        = {587--595}
}

@article{Qiu2018,
  author       = {Qiu, S. Q. and Dorrius, M. D. and de Jongh, S. J. and others},
  title        = {Micro-computed tomography (micro-CT) for intraoperative surgical margin assessment of breast cancer: a feasibility study in breast conserving surgery},
  journal      = {European Journal of Surgical Oncology},
  year         = {2018},
  volume       = {44},
  number       = {11},
  pages        = {1708--1713}
}

@article{Tang2013,
  author       = {Tang, R. and Coopey, S. B. and Buckley, J. M. and others},
  title        = {A pilot study evaluating shaved cavity margins with micro-computed tomography: a novel method for predicting lumpectomy margin status intraoperatively},
  journal      = {The Breast Journal},
  year         = {2013},
  volume       = {19},
  number       = {5},
  pages        = {485--489}
}

@article{Golshan2014,
  author    = {Golshan, M. and Sagara, Y. and Wexelman, B. and others},
  title     = {Pilot study to evaluate feasibility of image-guided breast-conserving therapy in the advanced multimodal image-guided operating (AMIGO) suite},
  journal   = {Annals of Surgical Oncology},
  year      = {2014},
  volume    = {22},
  pages     = {3356--3357}
}

@article{Papa2016,
  author    = {Papa, M. and Allweis, T. and Karni, T. and others},
  title     = {An intraoperative MRI system for margin assessment in breast conserving surgery: initial results from a novel technique},
  journal   = {Journal of Surgical Oncology},
  year      = {2016},
  volume    = {114},
  number    = {1},
  pages     = {22--26}
}

@article{Rossou2024,
  author    = {Rossou, E. and others},
  title     = {Reducing re-excision rates in breast conserving surgery with MarginProbe: systematic review},
  journal   = {British Journal of Surgery},
  year      = {2024},
  volume    = {111},
  number    = {1},
  doi       = {10.1093/bjs/znad335},
  url       = {https://academic.oup.com/bjs/article/111/1/znad335/7441103}
}

@misc{MarginProbeIFU2025,
  author       = {{Dilon Technologies}},
  title        = {MarginProbe Instructions for Use (Rev G)},
  year         = {2025},
  howpublished = {https://dilon.com/wp-content/uploads/2025/08/PB0501051-Rev-G-MarginProbe-Instruction-for-Use-IFU-US-for-print.pdf},

  note         = {Accessed August 2025}
}

@article{Kuwahara2020,
  author    = {Yoshihiko Kuwahara and Akira Nozaki and Kimihito Fujii},
  title     = {Large Scale Analysis of Complex Permittivity of Breast Cancer in Microwave Band},
  journal   = {Advances in Breast Cancer Research},
  volume    = {9},
  pages     = {101--109},
  year      = {2020},
  doi       = {10.4236/abcr.2020.94008},
  url       = {https://www.scirp.org/pdf/abcr_2020091414530938.pdf}
}

@article{Lazebnik2007,
  author    = {M. Lazebnik and E. L. Madsen and G. R. Frank and T. J. Yankeelov and J. H. Booske and M. Okoniewski and S. C. Hagness},
  title     = {Highly Accurate Debye Models for Normal and Malignant Breast Tissue Dielectric Properties at Microwave Frequencies},
  journal   = {IEEE Microwave and Wireless Components Letters},
  volume    = {17},
  number    = {12},
  pages     = {822--824},
  year      = {2007},
  doi       = {10.1109/LMWC.2007.910465},
  url       = {https://ieeexplore.ieee.org/document/4385287}
}

@book{hecht2012optics,
  title     = {Optics},
  author    = {Hecht, Eugene},
  year      = {2012},
  publisher = {Pearson},
  edition   = {4th}
}

@article{Sloane1979,
author = {Neil J. Sloane},
title = {Multiplexing Methods in Spectroscopy},
journal = {Mathematics Magazine},
volume = {52},
number = {2},
pages = {71-80},
year = {1979},
publisher = {Taylor & Francis},
doi = {10.1080/0025570X.1979.11976757},
URL = { 
        https://doi.org/10.1080/0025570X.1979.11976757
},
}

@book{harwit2012hadamard,
  title={Hadamard transform optics},
  author={Harwit, Martin},
  year={2012},
  publisher={Elsevier}
}

@article{Penketh2025,
  author       = {Penketh, Harry and Gallagher, Cameron P. and Mrnka, Michal and Lawrence, Christopher R. and Phillips, David B. and Hooper, Ian R. and Hendry, Euan},
  title        = {Hyperspectral imaging of microwave metasurfaces with deeply subwavelength resolution},
  journal      = {Nature Communications},
  year         = {2025},
  volume       = {16},
  number       = {1},
  pages        = {4612},
  doi          = {10.1038/s41467-025-59814-y},
  url          = {https://doi.org/10.1038/s41467-025-59814-y},
  publisher    = {Springer Nature}
}

@article{Lazebnik2007Meas,
  author       = {Mariya Lazebnik and Leah McCartney and Dijana Popovic and Cynthia B. Watkins and Mary J. Lindstrom and Josephine Harter and Sarah Sewall and Anthony Magliocco and John H. Booske and Michal Okoniewski},
  title        = {A large-scale study of the ultrawideband microwave dielectric properties of normal breast tissue obtained from reduction surgeries},
  journal      = {Physics in Medicine \& Biology},
  volume       = {52},
  number       = {10},
  pages        = {2637--2656},
  year         = {2007},
  publisher    = {IOP Publishing},
  doi          = {10.1088/0031-9155/52/10/001},
  url          = {https://doi.org/10.1088/0031-9155/52/10/001}
}

@article{Lazebnik2007MeasCancer,
  author       = {Mariya Lazebnik and Dijana Popovic and Leah McCartney and Cynthia B. Watkins and Mary J. Lindstrom and Josephine Harter and Sarah Sewall and Travis Ogilvie and Anthony Magliocco and Tara M. Breslin},
  title        = {A large-scale study of the ultrawideband microwave dielectric properties of normal, benign and malignant breast tissues obtained from cancer surgeries},
  journal      = {Physics in Medicine \& Biology},
  volume       = {52},
  number       = {20},
  pages        = {6093--6115},
  year         = {2007},
  publisher    = {IOP Publishing},
  doi          = {10.1088/0031-9155/52/20/002},
  url          = {https://doi.org/10.1088/0031-9155/52/20/002}
}

@article{Lazebnik2005,
  author       = {Mariya Lazebnik and Ernest L. Madsen and Gary R. Frank and Susan C. Hagness},
  title        = {Tissue-mimicking phantom materials for narrowband and ultrawideband microwave applications},
  journal      = {Physics in Medicine \& Biology},
  volume       = {50},
  number       = {18},
  pages        = {4245--4258},
  year         = {2005},
  publisher    = {IOP Publishing},
  doi          = {10.1088/0031-9155/50/18/001},
  url          = {https://doi.org/10.1088/0031-9155/50/18/001}
}

@article{Bray2024,
author = {Bray, Freddie and Laversanne, Mathieu and Sung, Hyuna and Ferlay, Jacques and Siegel, Rebecca L. and Soerjomataram, Isabelle and Jemal, Ahmedin},
title = {Global cancer statistics 2022: GLOBOCAN estimates of incidence and mortality worldwide for 36 cancers in 185 countries},
journal = {CA: A Cancer Journal for Clinicians},
volume = {74},
number = {3},
pages = {229-263},
doi = {https://doi.org/10.3322/caac.21834},
url = {https://acsjournals.onlinelibrary.wiley.com/doi/abs/10.3322/caac.21834},
year = {2024}
}

@Article{Yarina2022,
AUTHOR = {Yanina, Irina Y. and Nikolaev, Viktor V. and Zakharova, Olga A. and Borisov, Alexei V. and Dvoretskiy, Konstantin N. and Berezin, Kirill V. and Kochubey, Vyacheslav I. and Kistenev, Yuri V. and Tuchin, Valery V.},
TITLE = {Measurement and Modeling of the Optical Properties of Adipose Tissue in the Terahertz Range: Aspects of Disease Diagnosis},
JOURNAL = {Diagnostics},
VOLUME = {12},
YEAR = {2022},
NUMBER = {10},
ARTICLE-NUMBER = {2395},
URL = {https://www.mdpi.com/2075-4418/12/10/2395},
PubMedID = {36292084},
ISSN = {2075-4418},
DOI = {10.3390/diagnostics12102395}
}

@article{Saxena2024,
  author    = {Sonal Saxena and Ciaran Bench and Diksha Garg and Patric Boardman and Michal Mrnka and Harry Penketh and Nicholas Stone and Euan Hendry},
  title     = {Limitations of effective medium models for tissue phantoms in the THz frequency range},
  journal   = {Scientific Reports},
  volume    = {14},
  pages     = {22968},
  year      = {2024},
  doi       = {10.1038/s41598-024-70590-5},
  url       = {https://doi.org/10.1038/s41598-024-70590-5}
}

@article{Grant2024,
  author    = {N. E. Grant and S. L. Pain and E. Khorani and R. Jefferies and 
               A. Wratten and S. McNab and D. Walker and Y. Han and 
               R. Beanland and R. S. Bonilla and J. D. Murphy},
  title     = {Activation of Al$_2$O$_3$ surface passivation of silicon: Separating bulk and surface effects},
  journal   = {Applied Surface Science},
  volume    = {645},
  pages     = {158786},
  year      = {2024},
}

@article{Nicolson1970,
  author    = {Nicolson, A. M. and Ross, G. F.},
  title     = {Measurement of the intrinsic properties of materials by time-domain techniques},
  journal   = {IEEE Transactions on Instrumentation and Measurement},
  volume    = {IM-19},
  number    = {4},
  pages     = {377--382},
  year      = {1970},
  doi       = {10.1109/TIM.1970.4313932}
}

@article{Weir1974,
  author    = {Weir, W. B.},
  title     = {Automatic measurement of complex dielectric constant and permeability at microwave frequencies},
  journal   = {Proceedings of the IEEE},
  volume    = {62},
  number    = {1},
  pages     = {33--36},
  year      = {1974},
  doi       = {10.1109/PROC.1974.9382}
}
\end{document}